\documentclass[twocolumn]{aastex61}
\usepackage{natbib}
\usepackage{amsmath,bm}
\usepackage{subfigure}
\newcommand{\br}{{\mathbf{r}}}

\begin{document}

\title{Validating Forward Modeling and Inversions of Helioseismic Holography Measurements}

\author{K.~DeGrave}
\affiliation{NorthWest Research Associates, 3380 Mitchell Lane, Boulder, CO 80301, USA}

\author{D.~C.~Braun}
\affiliation{NorthWest Research Associates, 3380 Mitchell Lane, Boulder, CO 80301, USA}

\author{A.~C.~Birch}
\affiliation{Max-Planck-Institut f\"{u}r Sonnensystemforschung, Justus-von-Liebig-Weg 3, 37077 G\"{o}ttingen, Germany}

\author{A.~D.~Crouch}
\affiliation{NorthWest Research Associates, 3380 Mitchell Lane, Boulder, CO 80301, USA}

\author{B.~Javornik}
\affiliation{NorthWest Research Associates, 3380 Mitchell Lane, Boulder, CO 80301, USA}

\correspondingauthor{D.~C.~Braun}
\email{dbraun@nwra.com}

\begin{abstract}
Here we use synthetic data to explore the performance of forward models and inverse 
methods for helioseismic holography.  
Specifically, this work presents the first comprehensive test of inverse modeling for 
flows using lateral-vantage (deep-focus) holography.
We derive sensitivity functions in the Born 
approximation.  We then use these sensitivity functions in a series of forward 
models and inversions of flows from a publicly available magnetohydrodynamic 
quiet-Sun simulation.
The forward travel times computed using the kernels 
generally compare favorably with measurements obtained by applying holography, in a 
lateral-vantage configuration, on a 15-hour time series of artificial 
Dopplergrams extracted from the simulation.  Inversions for the horizontal 
flow components are able to reproduce the flows in the upper 3Mm of the domain, but 
are compromised by noise at greater depths. 

\end{abstract}

\section{Introduction}

Helioseismology has been a useful tool for studying the subsurface properties of 
the Sun. It is generally divided into global \citep[e.g.][]{JCD2003,Howe2009} and local 
\citep[e.g.][]{Gizon2010b,Braun2015} applications.  The former include inferences 
of the radially symmetric structure of the Sun and the latitudinal and
depth dependence of its internal rotation, while the latter include studies of 
the relatively small-scale structure and flows below sunspots, active regions 
and supergranulation, as well as larger-scale convection and
meridional circulation.  Giving us the means to 
indirectly image the sun's interior, helioseismology is of great importance to 
the study of solar structure and subsurface dynamics.

The forward problem in helioseismology is to determine the relationship between 
some quantity, measurable at the solar surface and an unobserved subsurface 
feature that we would like to study. For local applications like those 
considered in this work, 
the former are helioseismic wave travel-time measurements, and the latter are 
vector plasma flows. The two are related through linear integral equations of 
the form
\begin{equation}
  \delta \tau_{a}(\mathrm{\mathbf{r}}) = \int_\odot 
\sum_{\beta}K^{a}_{\beta}(\mathrm{\mathbf{r'}}-\mathrm{\mathbf{r}};z)
 v_{\beta}(\mathrm{\mathbf{r'}},z) 
\,\mathrm{d}^2\mathrm{\mathbf{r'}}\,\mathrm{d}z + n^{a}(\mathrm{\mathbf{r}}),
\end{equation}
\noindent where, for each travel-time measurement $\delta \tau_{a}$, $K^{a}_{\beta}$ are the three 
$\beta\in\{x,y,z\}$ 
components of a set of vector-valued functions called sensitivity kernels
and $n^{a}$ is the noise in these 
measurements. Here, $\mathbf{r} = (x,y)$ is the horizontal position and $z$ is the height 
(noting that $z=0$ at the surface
and $z < 0$ inside the Sun). Given these quantities, the inverse problem is to solve for 
the subsurface flow $v_{\beta}$ as accurately as possible through a series 
of matrix inversions.

In the last decade, the development and availability of realistic artificial data,
obtained from numerical wave-propagation computations, has allowed the validation and testing
of helioseismic procedures. Relevant numerical simulations include those 
computed under hydrostatic or magnetohydrostatic conditions
\citep[e.g.][]{Hanasoge2007b,Parchevsky2007,Hartlep2008,Felipe2010a} 
as well as fully compressible hydrodynamic or magnetohydrodynamic 
computations \citep[e.g.][]{Rempel2009,Rempel2015,Stein2009,Stein2012}.
Simulations give us the opportunity to test kernels and inversion 
procedures on data whose flow structure is known \textit{a priori}, and whose 
properties resemble the real Sun as closely as possible. This kind
of validation has been performed for both time-distance 
\citep[e.g.][]{Zhao2007,Svanda2011,DeGrave2014a,DeGrave2014b,Parchevsky2014} 
and helioseismic-holography 
methods \citep[e.g.][]{Braun2007,Birch2009,Braun2012,Dombroski2013,Braun2014}.
Validation tests of this kind are an important and necessary step in the 
helioseismic analysis of solar subsurface structure, as we are interested in 
recovering information from regions of the Sun that cannot be directly 
observed. 

In this work, we test the performance of helioseismic holography and kernels through a 
series of forward and inverse modeling comparisons, employing a realistic 
numerical simulation of the quiet Sun. 
The forward and inverse modeling tests we conduct employ a set of 
kernels computed for travel-times measured using helioseismic holography
(hereafter HH) carried out in a lateral-vantage geometry \citep{Lindsey2004b}. 
Until now, tests of lateral-vantage HH 
have only been carried out through comparisons of 
measured and forward-modeled travel times \citep[][]{Braun2007}. 
This work presents the first comprehensive test of inverse modeling for flows using
lateral-vantage HH.

The layout of the paper proceeds as follows: 
in \S\ref{sec.data} we describe the quiet-Sun simulation data used in this work. 
The holography travel-time 
measurement procedure is discussed in \S\ref{sec.ttmeas}, and the forward modeling 
is detailed in \S\ref{sec.forward}. 
The inversion method and results 
are described in \S\ref{sec.sola} and \S\ref{sec.invs}, and concluding remarks are 
given in \S\ref{sec.discuss}.

\section{Artificial Data}\label{sec.data}

\begin{figure}[htb]
\begin{center}$
\begin{array}{c}
\includegraphics[width=1.0\linewidth,clip=]{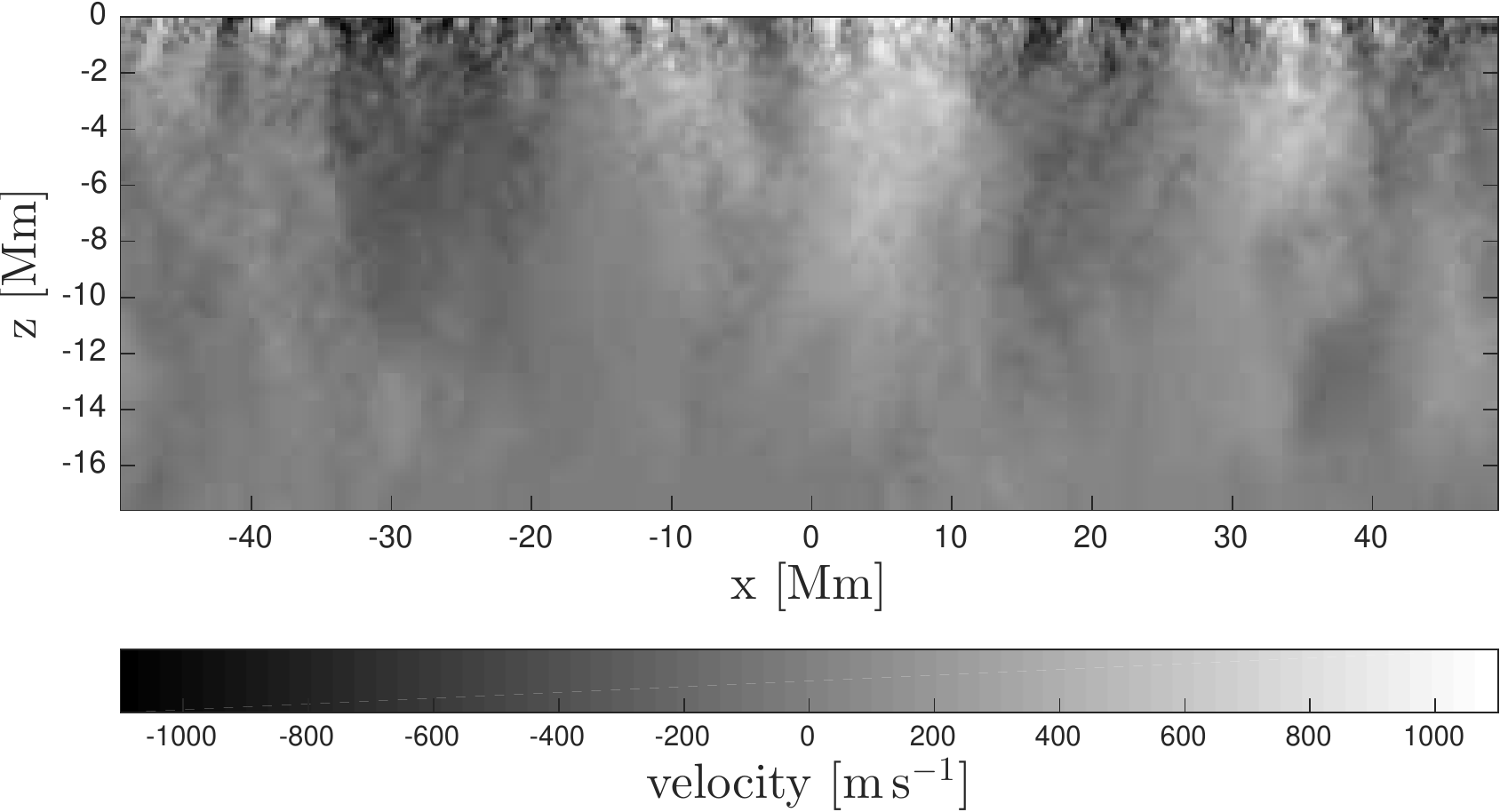}
\end{array}$
\end{center}
\caption{An example cut in depth through the $v_x$ flow component of the quiet-Sun 
simulation. The flows shown here have been averaged over $15$~hours.}
\label{simcut}
\end{figure}

The simulation we employ in this work represents quiet-Sun convection with a 
small-scale dynamo,
and has been described in detail in \citet{Rempel2014b}. 
A time-series of artificial data from this simulation
has been previously employed in validating the inversion procedures used in this
work \citep{DeGrave2014a}, using time-distance measurements and sensitivity 
kernels.
The simulation was computed with a horizontal and vertical resolutions
of 64 km and  32 km respectively. Convective motions 
excite surface gravity and acoustic waves which propagate 
throughout the domain which spans $98.3 \times 98.3$ Mm horizontally 
and 18.4 Mm vertically. A cut through the 
(15-hour) time-averaged $v_x$ flow component of the simulation is shown in 
Figure~\ref{simcut}. 

Doppler velocity 
time series, assuming a vertical line-of-sight, extracted at an optical 
depth of 0.01 
and having a cadence of $45$~seconds, are publicly 
available\footnote{\url{http://download.hao.ucar.edu/pub/rempel/sunspot\_models}}.
These artificial Dopplergrams have been interpolated from the original 
simulation onto a coarser grid with a horizontal spacing of 384 km.
For this work we utilize the first $15$~hr of the 
30-hr simulation run.

\section{Holography}\label{sec.ttmeas}

\begin{figure}[htb]
\epsscale{1.0}
\plotone{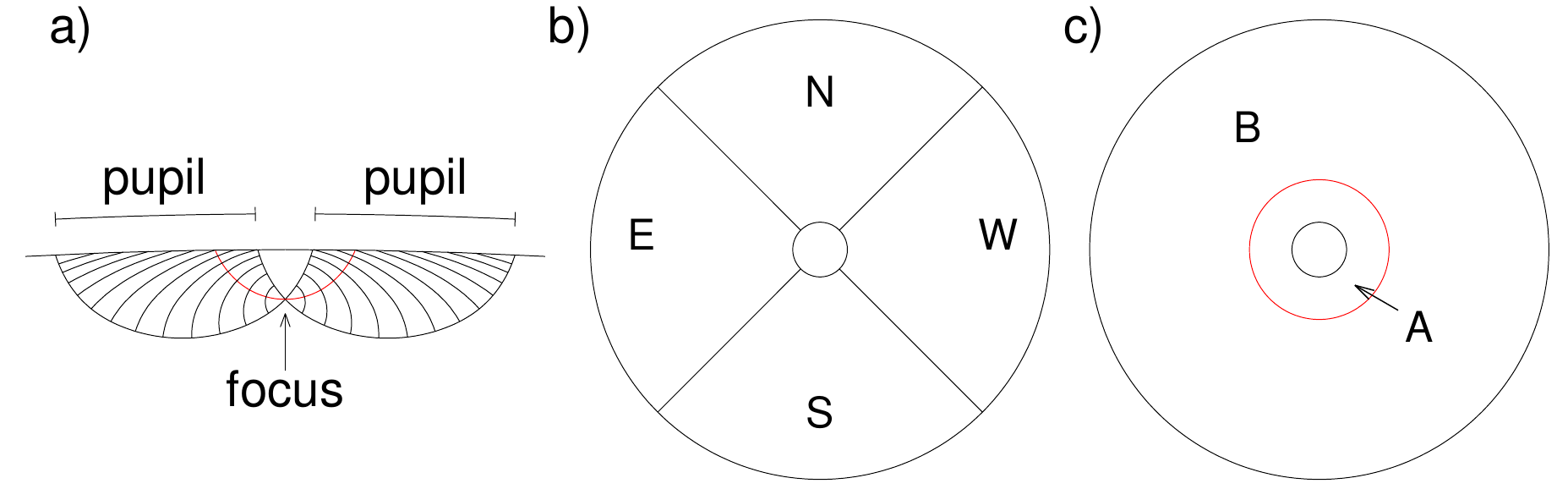}
\caption{
(a) side view and (b) top view of the pupil quadrants employed in 
lateral-vantage HH.
The ray path colored in red in panel (a) corresponds to waves propagating 
horizontally
through the focus.
Panel (c) shows the full annulus divided into inner and outer pupils (labeled A 
and B respectively)
separated by the red circle which represents the intersection of the surface 
with the 
ray paths shown in red in panel (a). 
Example loci of constant phase are
shown for the egression/ingression amplitudes in panel (a).
}
\label{fig.lv_geom}
\end{figure}

Helioseismic holography is a method which
computationally extrapolates the surface acoustic field
from a selected pupil into the solar interior \citep{Lindsey1997}
in order to estimate the complex amplitudes of the waves propagating into
or out of a focus point at a chosen depth within the solar
interior. These amplitudes are called the acoustic ingression and
acoustic egression respectively. 
Lateral-vantage holography \citep[e.g.][]{Lindsey2004b} is analogous to deep-focus
methods in time-distance helioseismology and common-depth-point reflection
terrestrial seismology. 
Figure~\ref{fig.lv_geom} illustrates the pupil geometry used. The annulus
is defined by rays
propagating through the focus and inclined up to ${\pm 45}^\circ$ from the direction
parallel to the surface. 
The practical aspects of the methodology have been described in detail elsewhere
\citep[e.g.][]{Braun2007,Braun2008b,Braun2014}. Most of the prior applications
of lateral-vantage HH make use of
the northward-minus-southward (NS) and westward-minus-eastward
(WE) travel-time differences $\delta \tau_\mathrm{ns}$ and $\delta \tau_\mathrm{we}$ 
as derived from cross-covariances between the egression and
ingression as assessed using opposite quadrants (Figure~\ref{fig.lv_geom}b).
Here, an additional pair of cross-covariances are obtained
using inner and outer portions of the complete annulus 
(Figure~\ref{fig.lv_geom}c).
The radius $\rho_\mathrm{h}$ which separates the two subannuli is defined by the 
ray path of a wave propagating horizontally through the focus.
Cross-covariances between egressions and ingressions assessed in these
subannuli are used to determine an outward-minus-inward (OI) propagation
travel-time difference $\delta \tau_\mathrm{oi}$.
As described elsewhere \citep{Braun2014}, Gaussian phase-speed filters are also used, 
with a width $\delta w$ and a peak at $w_\mathrm{o}$ corresponding to the phase-speed of the 
aforementioned horizontally propagating wave. The minimum and maximum radii of the annulus, 
${\rho}_\mathrm{min}$ and  ${\rho}_\mathrm{max}$, horizontal-ray radius $\rho_\mathrm{h}$, and phase-speed
filter parameters for the measurements used here are given by Table~\ref{tbl-lv}. 

\begin{deluxetable}{cccccc}
\tablecolumns{6}
\tablewidth{0pc}
\tablecaption{Lateral-vantage: pupil size and filter parameters\label{tbl-lv}}
\tablehead{
\colhead{focus depth} & \colhead{${\rho}_\mathrm{min}$} & \colhead{${\rho}_\mathrm{h}$} & 
\colhead{${\rho}_\mathrm{max}$} & \colhead{$w_\mathrm{o}$} & \colhead{$\delta w$} \\ 
\colhead{(Mm)} & \colhead{(Mm)} & \colhead{(Mm)} & \colhead{(Mm)} & 
\colhead{(km s$^{-1}$)} & \colhead{(km s$^{-1}$)} \\ 
}
\startdata
      0.77  &       1.0 &       5.5 &       13.9 &   13.6 &        6.6 \\
      1.53  &       1.2 &       5.8 &       14.6 &   15.3 &        7.4 \\
      2.30  &       1.6 &       6.3 &       16.0 &   17.1 &        8.3 \\
      2.99  &       2.1 &       7.0 &       16.7 &   18.8 &        9.2 \\
      3.97  &       2.8 &       7.7 &       18.1 &   21.0 &       10.5 \\
      5.01  &       3.5 &       9.0 &       20.2 &   23.6 &      11.8  \\
      5.99  &       3.5 &       9.7 &       24.4 &   26.7 &      13.1  \\
      6.96  &       4.2 &      10.4 &       31.3 &   29.3 &      14.9  \\
      8.35  &       4.9 &      11.8 &       39.0 &   33.7 &      16.6  \\
\enddata
\end{deluxetable}

\section{Forward Modeling}\label{sec.forward}

To compute kernels for deep-focusing travel times measured from the synthetic observations, 
we first construct a model power spectrum by fitting the power spectrum of the synthetic data.
We fit for a source function (this determines the mode amplitude), damping rate, and deviation 
of the frequency from the Model~S \citep{JCD1996} frequency as functions
of horizontal wavenumber and radial order.  The details are described in 
Appendix~\ref{sec.model}.  
We then use the Born approximation
to compute the sensitivity of the lateral-vantage travel-time differences to small-amplitude 
steady flows.  The calculation is based on the approach of \citet{Birch2007a},
though extended to include the holography Green’s functions and the pupil functions. 
The product of the Green’s function and the pupils
appears in the calculation in exactly the same way as a non-axisymmetric complex-valued 
data analysis filter. In this calculation we use the source function and damping rates 
obtained from the fit to the power spectrum.  We, however, do not include in the calculations the 
effect of the changes in the mode frequencies but instead use the normal-mode frequencies and 
eigenfunctions of Model~S.  
We estimate that this approximation introduces an error of about 2\,\% in the kernels.
For the applications in this work, this error is significantly smaller than the noise
in the travel time measurements (see \S\ref{sec.forcomp}).
Appendix~\ref{sec.kernels}  shows the details of the calculation.

\subsection{Model for the Noise Covariance}\label{sec.ttnoise}

Helioseismic travel-time measurements contain random noise due to the stochastic nature 
of the convective forcing that drives solar oscillations.  This noise is important to 
characterize as it propagates through our inversions and ultimately gives rise to 
uncertainties in the recovered flows.  We estimate the level of noise in our 
measurements by computing the travel-time noise covariance matrix using 200 Monte-Carlo realizations
of stochastic wavefields following the procedure of \citet{Gizon2004b}. 

\subsection{Forward Comparisons}\label{sec.forcomp}

\begin{deluxetable}{ccccc}
\tablecolumns{5}
\tablewidth{0pc}
\tablecaption{Lateral-vantage correlation statistics comparing measured and 
forward-modeled travel-time differences.\label{tbl-lvfm}}
\tablehead{
\colhead{travel-time} & \colhead{focus depth}  & \colhead{RMS error} & 
\colhead{slope} \\
\colhead{difference} & \colhead{(Mm)}  & \colhead{(s)} & 
\colhead{}\\}
\startdata
$\delta\tau_\mathrm{we}$ & 0.77 & 7.3 & 1.02\\
$\delta\tau_\mathrm{we}$ & 1.53 & 5.6 & 0.99\\
$\delta\tau_\mathrm{we}$ & 2.30 & 4.7 & 1.03\\
$\delta\tau_\mathrm{we}$ & 2.99 & 4.3 & 1.00\\
$\delta\tau_\mathrm{we}$ & 3.97 & 3.8 & 0.98\\
$\delta\tau_\mathrm{we}$ & 5.01 & 3.8 & 0.98\\
$\delta\tau_\mathrm{we}$ & 5.99 & 3.8 & 0.98\\
$\delta\tau_\mathrm{we}$ & 6.96 & 3.8 & 0.98\\
$\delta\tau_\mathrm{we}$ & 8.35 & 4.0 & 0.94\\\\
$\delta\tau_\mathrm{oi}$ & 0.77 & 3.7 & 1.01\\
$\delta\tau_\mathrm{oi}$ & 1.53 & 3.3 & 0.98\\
$\delta\tau_\mathrm{oi}$ & 2.30 & 3.3 & 1.01\\
$\delta\tau_\mathrm{oi}$ & 2.99 & 3.2 & 1.00\\
$\delta\tau_\mathrm{oi}$ & 3.97 & 3.2 & 1.03\\
$\delta\tau_\mathrm{oi}$ & 5.01 & 3.4 & 1.06\\
$\delta\tau_\mathrm{oi}$ & 5.99 & 3.2 & 1.04\\
$\delta\tau_\mathrm{oi}$ & 6.96 & 3.2 & 1.01\\
$\delta\tau_\mathrm{oi}$ & 8.35 & 3.4 & 0.99\\
\enddata
\end{deluxetable}

\begin{figure*}[htb]
\begin{center}$
\begin{array}{c}
\includegraphics[width=0.9\linewidth,clip=]{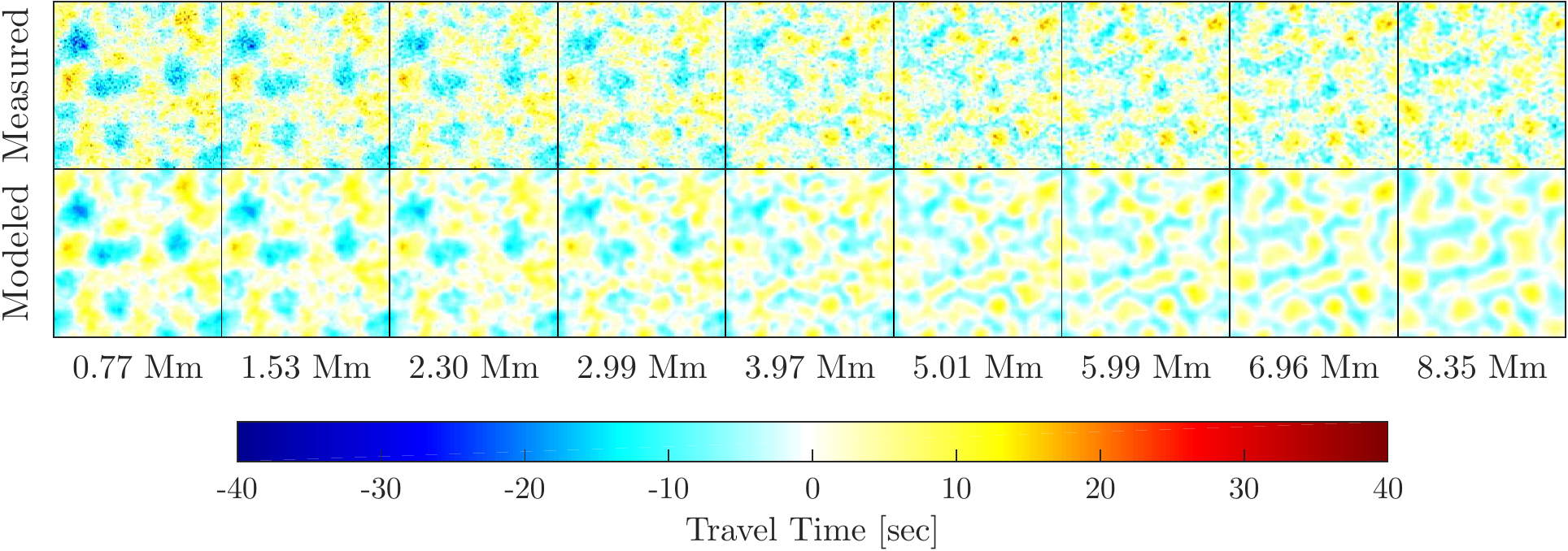}
\end{array}$
\end{center}
    \caption{Measured (top row) and forward-modeled (bottom row) OI 
lateral-vantage travel-time maps for each of the nine focus depths (columns).
The panels span the full $98.3 \times 98.3$ Mm horizontal range of the simulation.
The focus depth is shown at the bottom of each column.}
\label{LVtmaps}
\end{figure*}

Figure~\ref{LVtmaps} shows comparisons of travel-time difference maps measured 
from the simulation using HH with maps predicted from the sensitivity functions 
convolved with the true time-averaged flows present in the simulation. Correlation statistics 
comparing each pair of measured and forward-modeled maps are shown in 
Table~\ref{tbl-lvfm} for the WE and OI travel-time differences. The table 
includes the root-mean-square of the difference between maps (RMS error) and 
the slope of the least-squares linear fit between measured and modeled values 
(the fit assumes there are no errors in the modeled values). Good agreement is found between 
measured and forward-modeled travel-time maps, and slope values are close to 
unity for all focus depths. 

The RMS errors have values between 4--7 seconds for the WE measurements, and between 3--4 seconds for the OI
measurements.  For context, the forward-modeled maps exhibit travel-time differences with 
peak values ranging from 60 seconds for the shallowest focus depth to about 15 seconds 
for the deepest measurements. We note that a 2\,\% error in the kernels (as discussed above) would
produce travel-time errors which are, for the most part, considerably less than one second.
This is significantly smaller than the RMS errors.

\section{SOLA Inversion Method}\label{sec.sola}

To recover flows from the simulation, we employ the Subtractive Optimally 
Localized Averaging (SOLA) method \citep{Pijpers1992}. The goal of SOLA is to 
find a set of two-dimensional inversion weights 
\citep[see][]{Svanda2011,Jackiewicz2012} that, when spatially convolved with 
the travel-time measurements, will give a smoothed estimate of flow component 
$\alpha=\{x,y,z\}$ at some target depth $z_0$ within the simulation domain: 
\begin{equation}
  v^{\mathrm{inv}}_{\alpha}(\mathrm{\mathbf{r}};z_0)=\sum_{i}\sum^{M}_{a=1} w^{a}_{\alpha}(\mathrm{\mathbf{r}}_i-\mathrm{\mathbf{r}};z_0)\delta\tau_a(\mathrm{\mathbf{r}}_i),\label{wt}
\end{equation}
where $\delta\tau_a$ are the set of $M$ travel-time measurements, 
and $w^{a}_{\alpha}$ are their respective weights. 
The sum over $i$ is over all horizontal positions.
When a set of weights has 
been computed, they are linearly combined with the sensitivity kernels to 
produce a so-called averaging kernel:
\begin{equation}
  \mathcal{K}^{\beta}_{\alpha}(\mathrm{\mathbf{r}},z;z_0)=\sum_{i}\sum^{M}_{a=1} w^{a}_{\alpha}(\mathrm{\mathbf{r}}_i-\mathrm{\mathbf{r}};z_0){K}^{a}_{\beta}(\mathrm{\mathbf{r}}-\mathrm{\mathbf{r}}_i;z)
\label{wk}
\end{equation}
which effectively gives the spatial resolution of the inversion. Here, 
$\beta=\{x,y,z\}$ are the three kernel components, and the sum over $i$ 
represents a horizontal convolution. Ideally, the weights 
will be such that, for $\alpha = \beta$, the resulting averaging kernel is well-localized in 
three-dimensional space, closely matching a pre-defined (typically Gaussian) 
target function, $T$. 
For $\beta \neq \alpha$ the averaging kernel will ideally be small - these off-diagonal 
components are responsible for cross-talk \citep{Jackiewicz2012}.

\begin{figure*}[htb]
\begin{center}$
\begin{array}{ccc}
\includegraphics[width=0.3\linewidth,clip=]{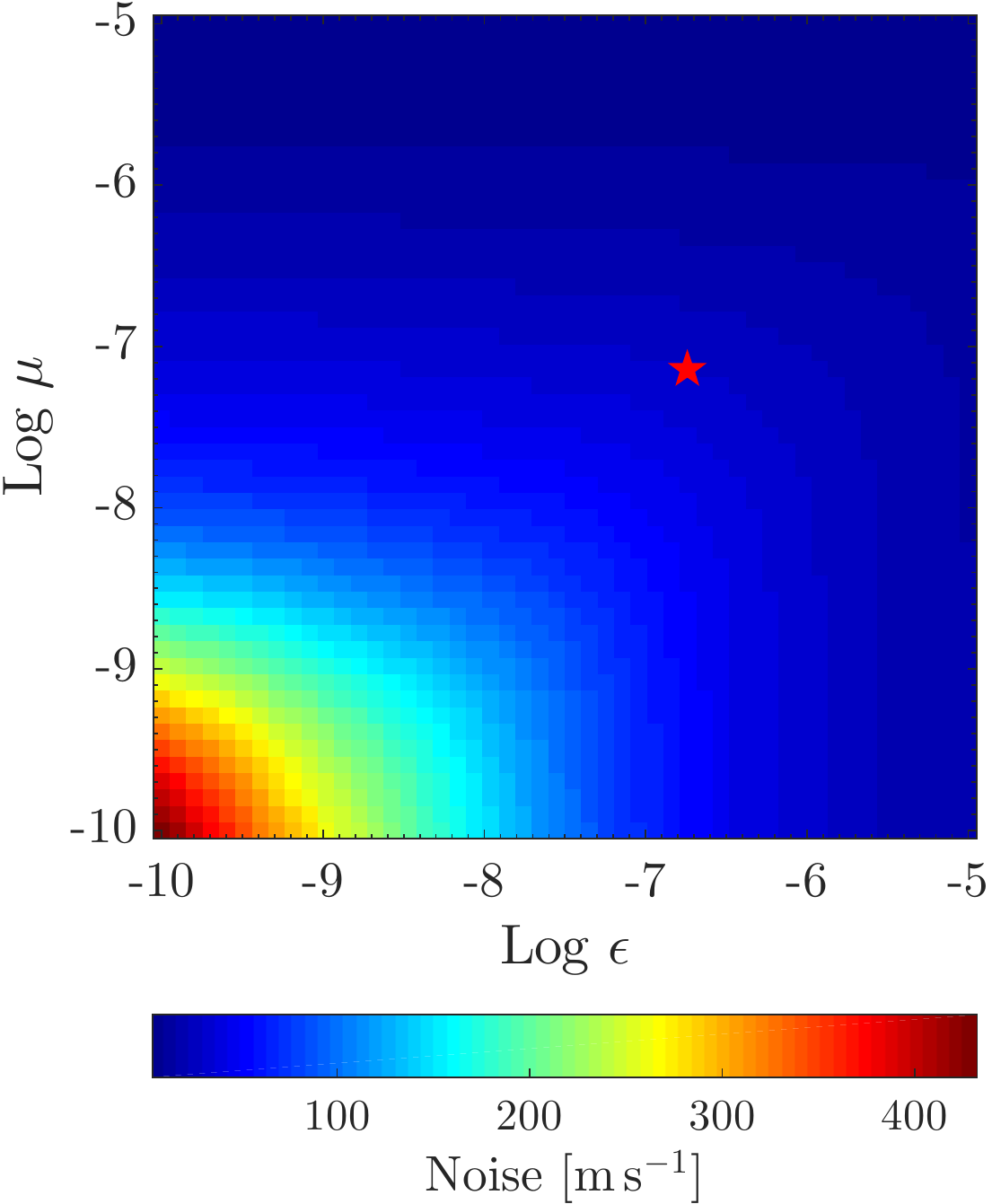}&
\includegraphics[width=0.3\linewidth,clip=]{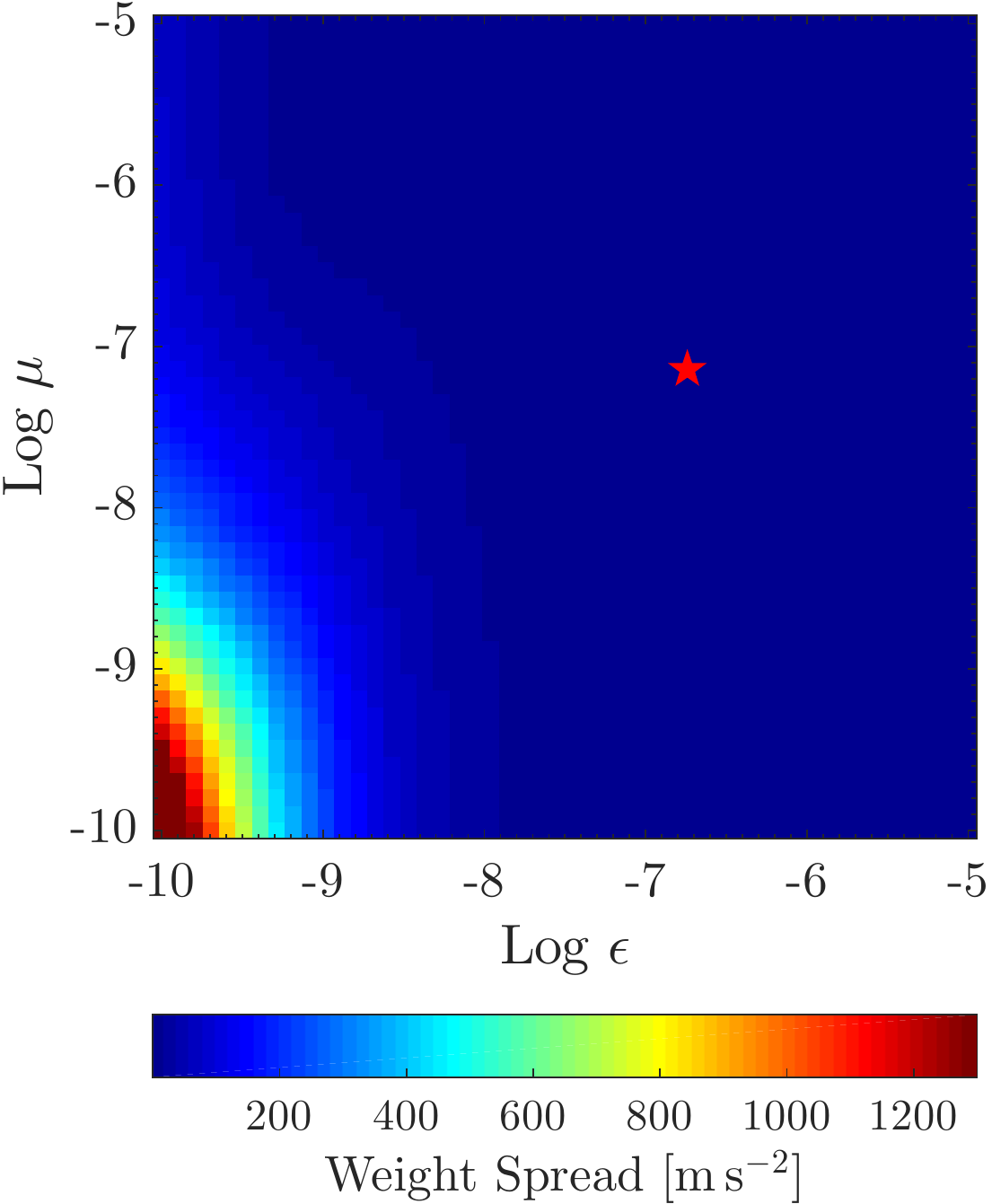}&
\includegraphics[width=0.3\linewidth,clip=]{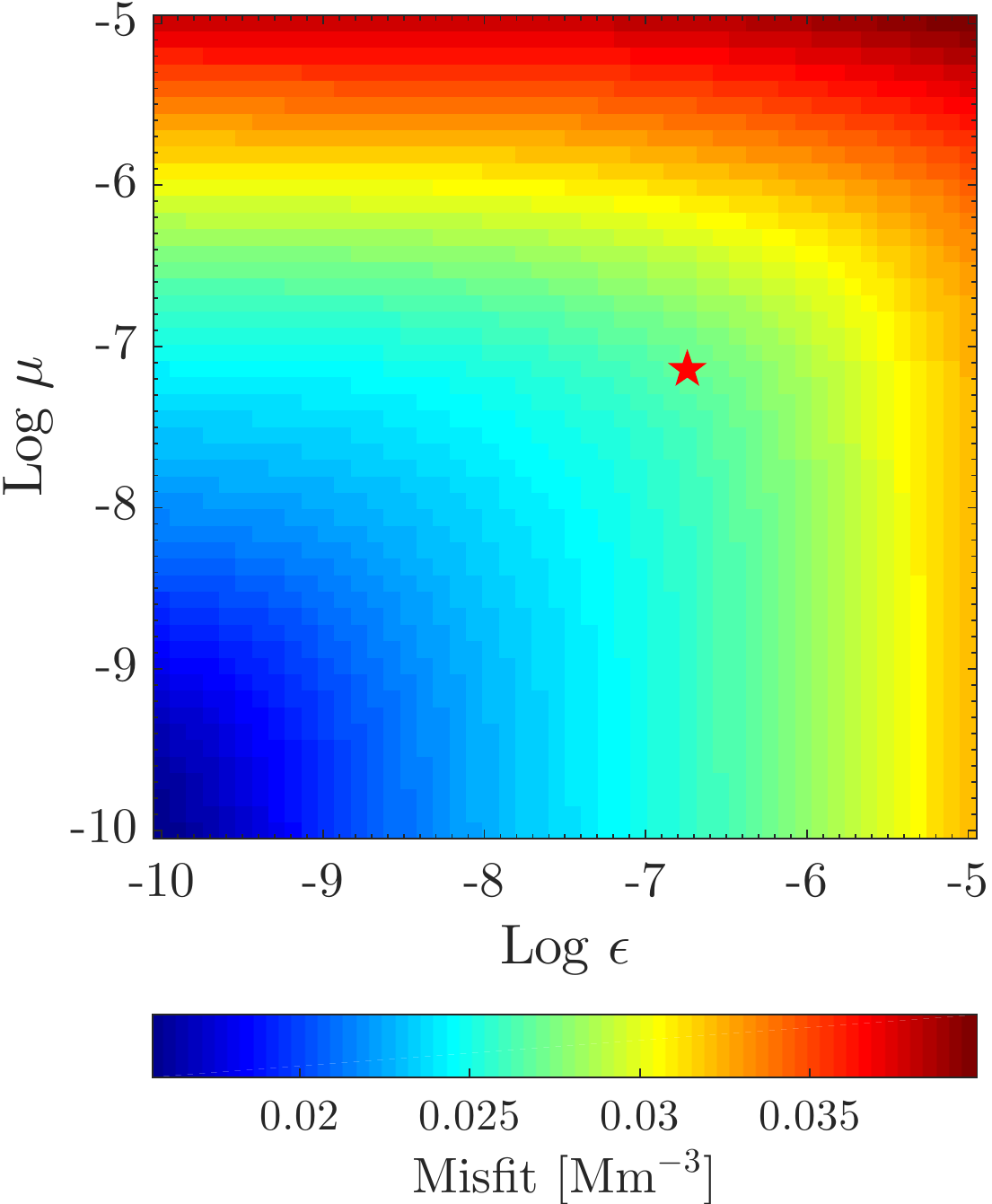}
\end{array}$
\end{center}
\caption{Example solution grids for the $1\,\rm{Mm}$ depth $v_x$ inversion 
showing inversion noise level (left), weight spread (middle), and misfit 
(right) values for every combination of regularization parameters $\mu$ and 
$\epsilon$. The cross-talk is not regularized in the horizontal flow 
inversions, and so $\nu$ has been set to zero. The red star denotes the 
solution that minimizes the misfit and weight spread for a $30\,\rm{m\,s^{-1}}$ 
noise level.}
\label{sgrid}
\end{figure*}

For each inversion target depth $z_0$ and target flow direction $\alpha$, we search for a set of weights that minimizes 
the cost function
\begin{subequations}
\begin{align}
  \mathcal{X}=\int_\odot 
[\mathcal{K}^{\alpha}_{\alpha}(\mathrm{\mathbf{r}},z;z_0)-T(\mathrm{\mathbf{r}},
z;z_0)]^2 \,\mathrm{d}^2\mathrm{\mathbf{r}}\,\mathrm{d}z \\ +\, 
\nu\sum_{\beta\neq\alpha}\int_\odot 
[\mathcal{K}^{\beta}_{\alpha}(\mathrm{\mathbf{r}},z;z_0)]^2 
\,\mathrm{d}^2\mathrm{\mathbf{r}}\,\mathrm{d}z \\ +\, 
\epsilon\sum_{a,i}[w^a_{\alpha}(\mathrm{\mathbf{r}}_i;z_0)]^2 \\ +\, 
\mu\sum_{a,b,i,j} 
w^a_{\alpha}(\mathrm{\mathbf{r}}_i;z_0)\Lambda_{ab}w^b_{\alpha}(\mathrm{\mathbf{
r}}_j;z_0),
\end{align}
\label{cost}
\end{subequations}

\noindent where the various terms in Equation~\ref{cost} represent the misfit 
between the averaging kernel and target function (\ref{cost}a); the extent to which 
the non-inverted-for flow components contribute to the recovered velocities, 
referred to as cross-talk (\ref{cost}b); an ad hoc term quantifying 
the localization of the inversion weights, referred to as the weight spread (\ref{cost}c); 
and the level of random 
noise in the solution (\ref{cost}d). $\Lambda_{ab}$ is the noise covariance.
These terms can be controlled to some degree by 
varying regularization parameters $\nu$, $\epsilon$, and $\mu$. In practice, we 
perform inversions and compute their respective \ref{cost}a\,--\,\ref{cost}d quantities for 
many combinations of regularization values for a given target depth. An
optimal combination of parameters is then selected from these. Example solution 
grids for a typical horizontal flow inversion are shown in Figure~\ref{sgrid}.

When performing inversions for the horizontal flow components ($v_x,v_y$) in 
this work, priority is first placed on choosing an acceptable noise level of 
roughly $30\,\rm{m\,s^{-1}}$ for each target depth. This noise level seems 
reasonable given the fact that flows in the upper $5\,\rm{Mm}$ of the 
simulation domain are of the order $300\,\rm{m\,s}^{-1}$. 
Of the various 
combinations of regularization parameters that yield the selected noise level, 
we choose the solution that provides a reasonable trade-off between the misfit 
and weight spread. This is 
done through a typical L-curve analysis.
For horizontal flow inversions, we set 
$\nu=0$, and therefore do not regularize the amount of cross-talk, as its 
effects are small compared to the large-amplitude flows that we are inverting 
for.

Inversions for the vertical flow component, $v_z$, also place priority on first 
selecting an acceptable level of noise. The chosen noise level must be much 
lower than that of the horizontal flow inversions, as the amplitude of the 
vertical flows in the top layers of the simulation domain are roughly an order 
of magnitude weaker than the horizontal flows ($\approx15\,\rm{m\,s^{-1}}$ on 
supergranule scales). We therefore choose a noise level of $5\,\rm{m\,s^{-1}}$ 
in order to retain the possibility of recovering the weak vertical flow signal 
while balancing the misfit between averaging kernel and target function.

Unlike the horizontal flow inversions, we also now fix $\nu\approx100$, in 
addition to varying the misfit and weight spread regularization parameters. 
Constraining cross-talk is necessary for vertical flow 
inversions; failing to do so can give rise to cross-talk which has the 
same amplitude or greater than the weak vertical flows for which we are 
inverting. This often leads to recovered flows that are strongly anticorrelated 
with the true flows \citep[e.g.,][]{Zhao2007}. Of the solutions that yield the 
specified noise level, we again choose the one that provides a reasonable 
trade-off between the misfit and weight spread.

\section{Inversion Results}\label{sec.invs}

The code used to perform inversions in this work has been 
validated previously with time-distance measurements applied to realistic 
simulations \citep{DeGrave2014a,DeGrave2014b}. 
To assess the performance of the inversions, we compare the recovered flows, 
$v^{\mathrm{inv}}_{\alpha}$, with 
the target flows, which represent the true flows
$v^{\mathrm{sim}}_{\alpha}$ 
smoothed to the expected resolution of the inversion flow maps via convolution 
with the inversion target function:
\begin{equation}
  v^{\mathrm{tgt}}_{\alpha}(\mathrm{\mathbf{r}},z_0)=\int_\odot
  T(\mathrm{\mathbf{r'}}-\mathrm{\mathbf{r}};z,z_0)v^{\mathrm{sim}}_{\alpha}(\mathrm{\mathbf{r'}},z) \,
  \mathrm{d}^2\mathrm{\mathbf{r'}}\,\mathrm{d}z.
\label{tv}
\end{equation}
The target solution represents the best we can hope to achieve in any 
particular inversion. 
For the vertical flow inversion, we also examine the cross-talk components, 
\begin{equation}
  v^{\beta}_{\alpha}(\mathrm{\mathbf{r}};z_0)=\int_\odot
\mathcal{K}^{\beta}_{\alpha}(\mathrm{\mathbf{r'}}-\mathrm{\mathbf{r}},z;z_0) 
v^{\mathrm{sim}}_{\beta}(\mathrm{\mathbf{r'}},z) \,
\mathrm{d}^2\mathrm{\mathbf{r'}}\,\mathrm{d}z
\label{kv}
\end{equation}
in addition to $v^{\mathrm{inv}}_{\alpha}$. We note that 
$v^{\mathrm{inv}}_{\alpha}\equiv v^{\alpha}_{\alpha}$ when the level of noise 
is negligible.

\subsection{Horizontal Flow Inversions}\label{sec.hinvs}

\begin{figure*}[htb]
\begin{center}$
\begin{array}{ccc}
\includegraphics[width=0.356\linewidth,clip=]{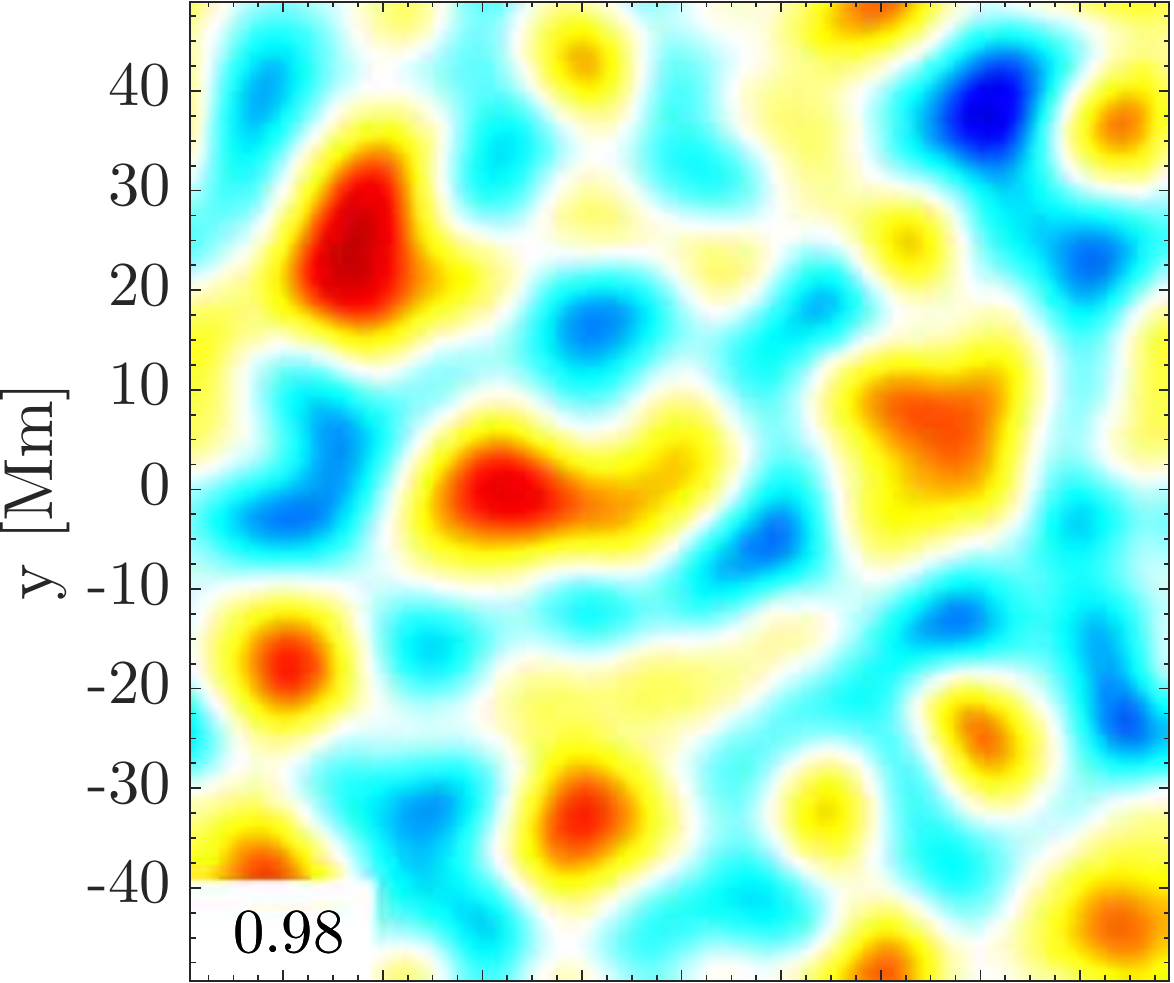}&
\includegraphics[width=0.3\linewidth,clip=]{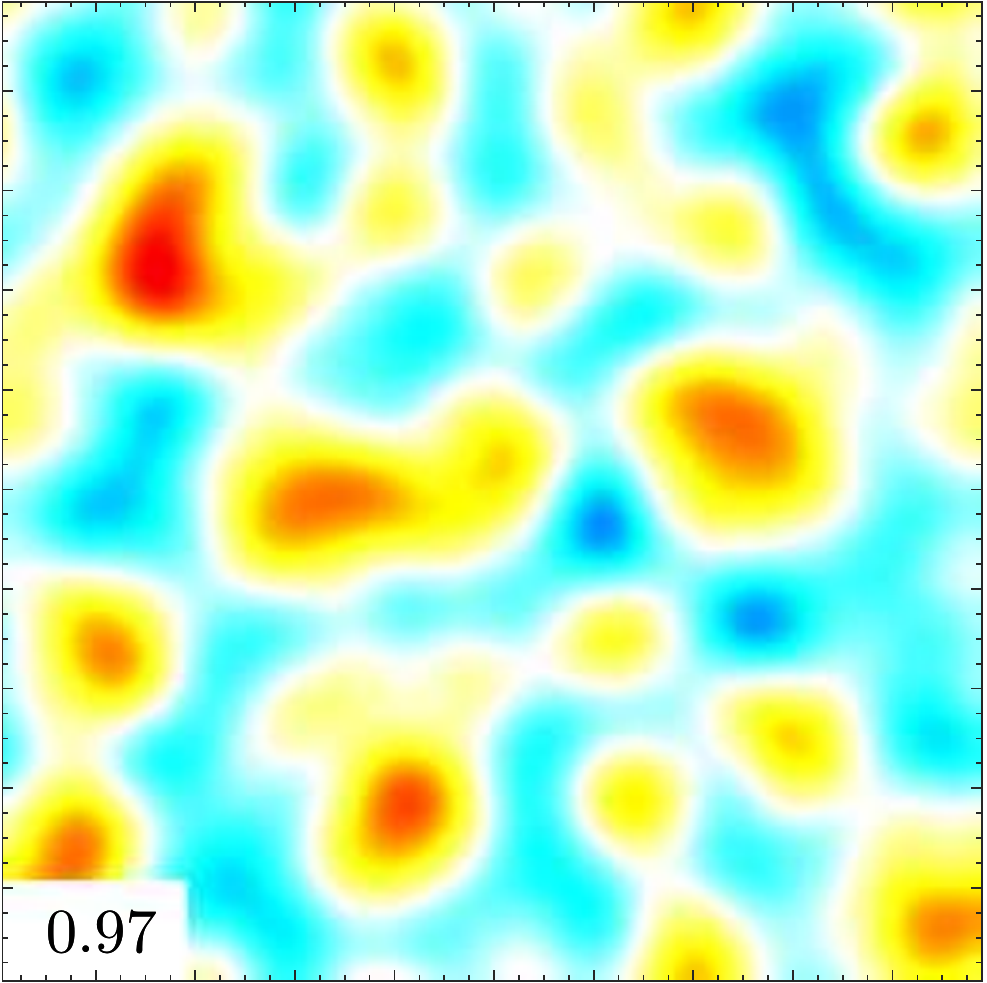}&\vspace{0.1cm}
\includegraphics[width=0.3\linewidth,clip=]{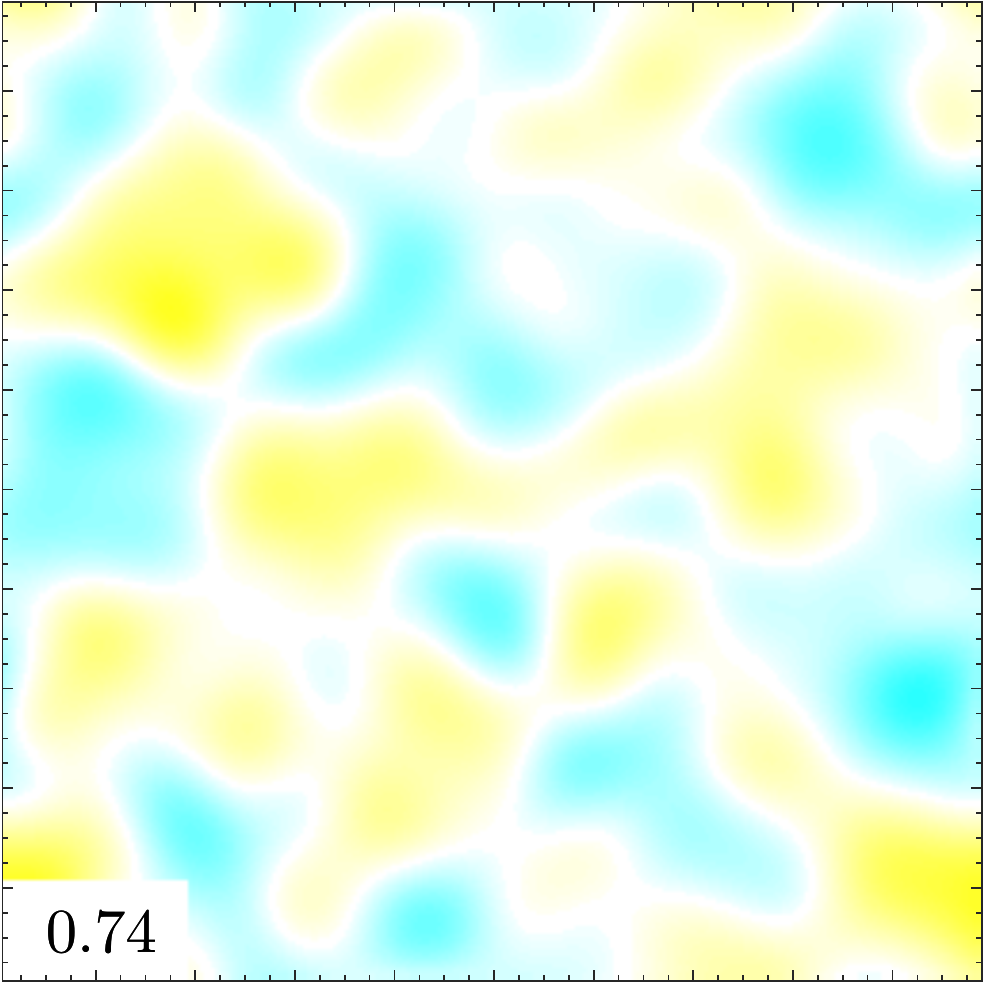}\\
\includegraphics[width=0.356\linewidth,clip=]{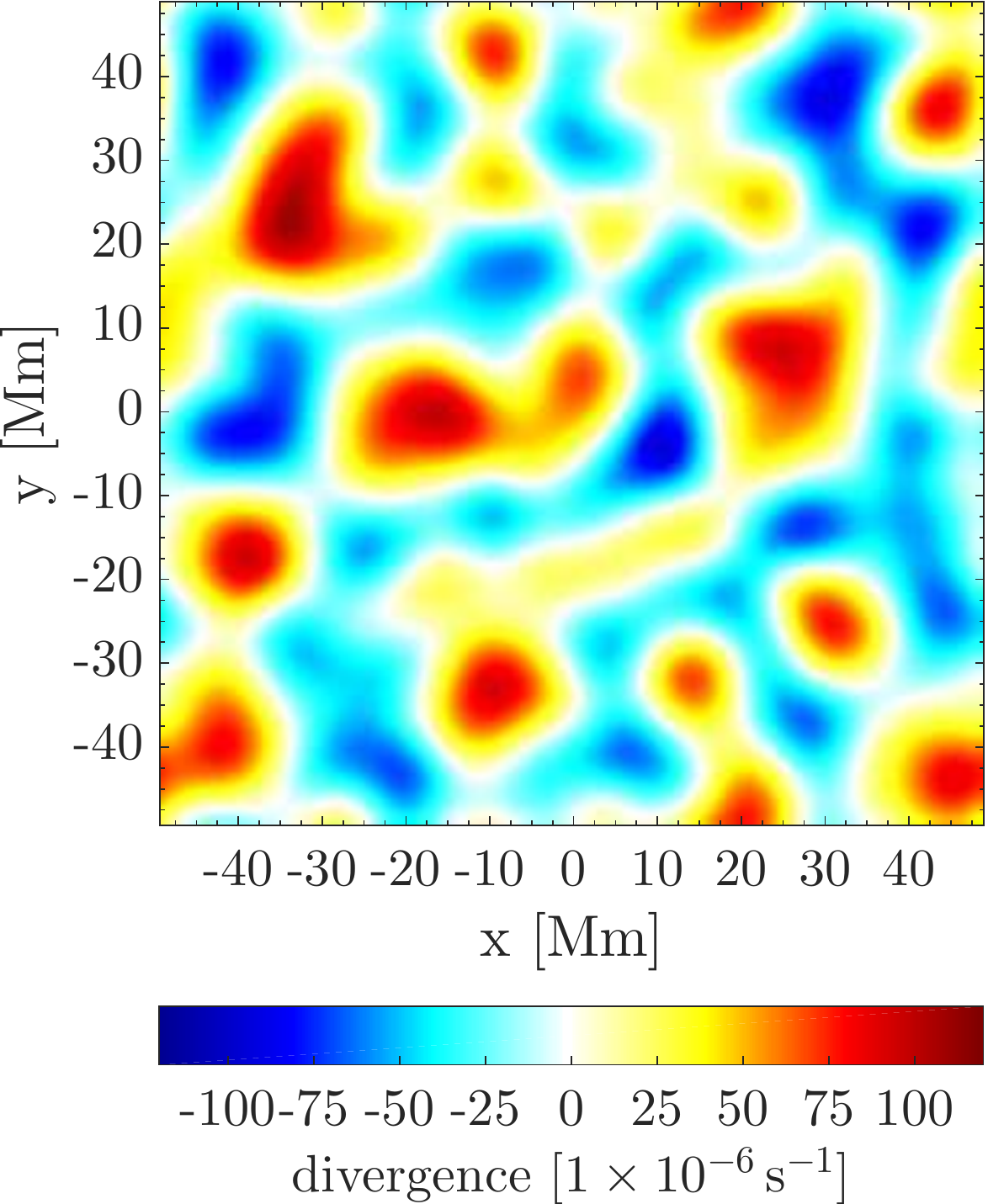}&
\includegraphics[width=0.3\linewidth,clip=]{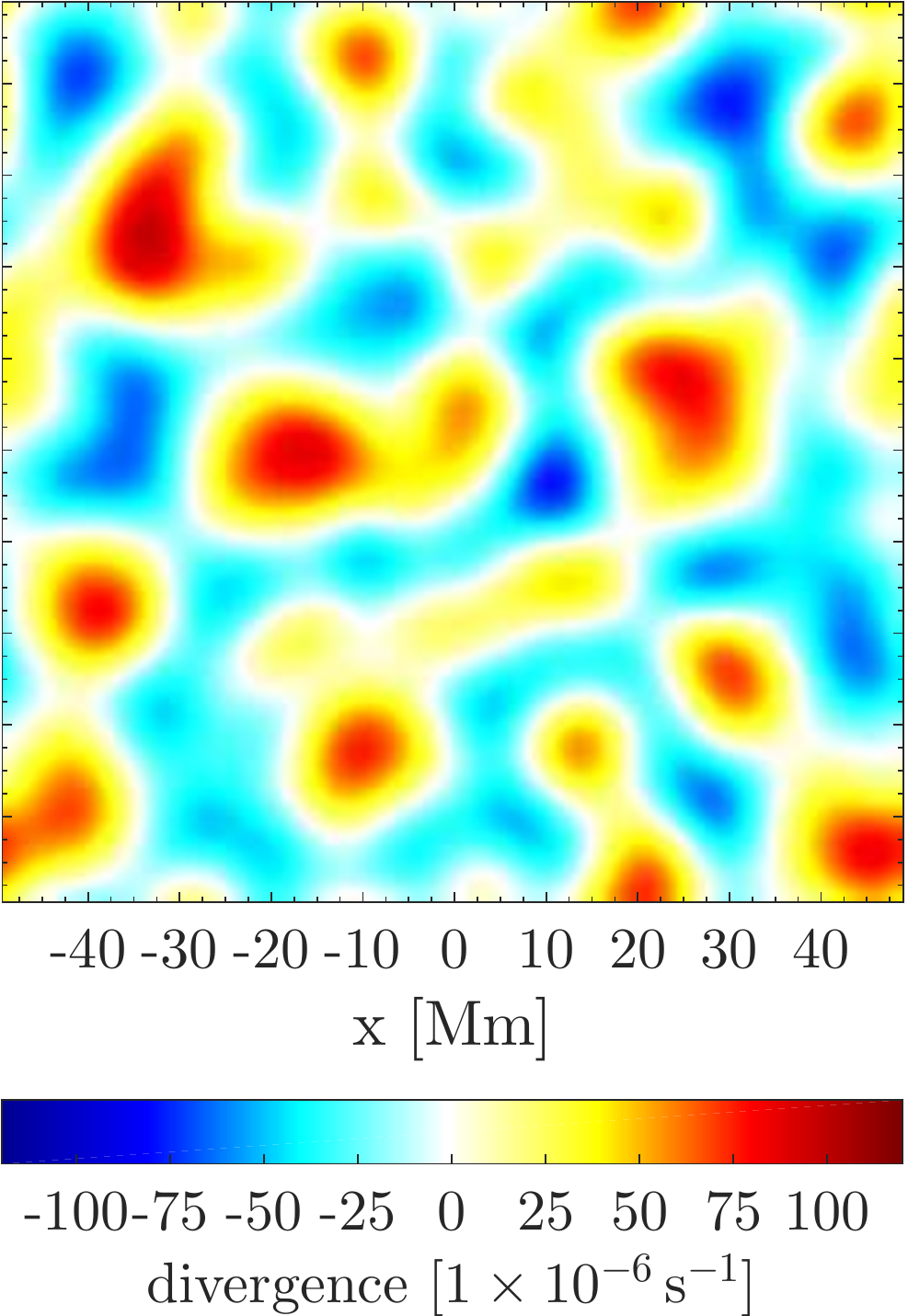}&
\includegraphics[width=0.3\linewidth,clip=]{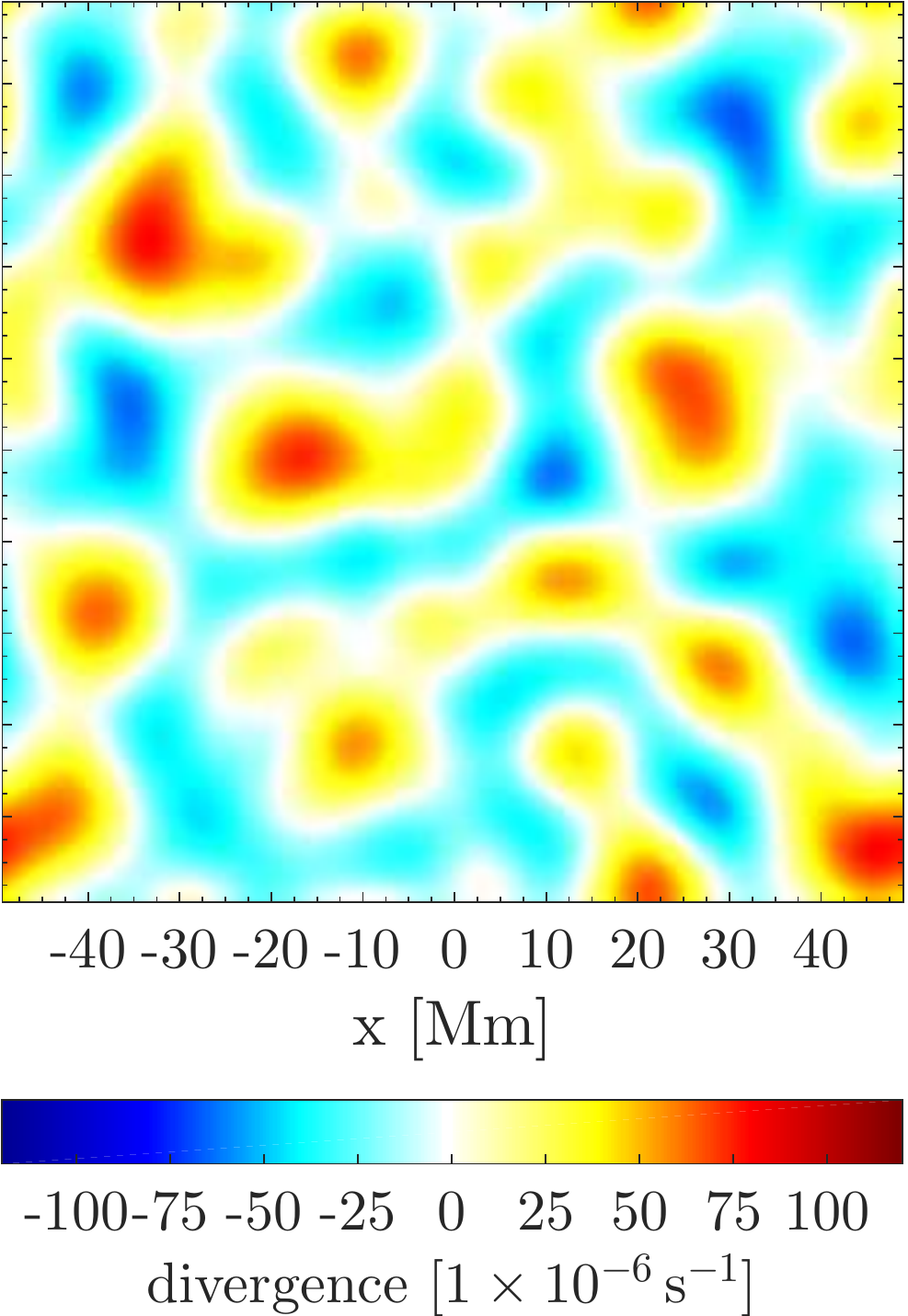}
\end{array}$
\end{center}
\caption{Maps showing the horizontal flow divergence derived from the 
($v_x,v_y$) flows recovered from the $1\,\rm{Mm}$ (left column), $3\,\rm{Mm}$ 
(middle column), and $5\,\rm{Mm
}$ (right column) depth inversions. The target simulation flows at these depths 
are shown in the bottom row. Correlation values between the inversion
and target simulation flows are shown in the bottom left-hand corner of each 
top-row panel. All maps share the same color scale.}
\label{qsvx}
\end{figure*}

Inversions for the horizontal flow components ($v_x,v_y$) were carried out at 
depths of $1$, $3$, and $5\,\rm{Mm}$ below the surface of the simulation 
domain. Figure~\ref{qsvx} shows maps of the resulting lateral-vantage 
inversions (top row), along with the smoothed simulation flows, 
$v^{\mathrm{tgt}}_{x}$, at each of these depths (bottom row). The 
two-dimensional Pearson correlation values between the inversion and target 
simulation flows are given in the lower left-hand corner of each panel. The 
noise level for each of the inversions is approximately $30\,\rm{m\,s^{-1}}$ 
for all target depths. The horizontal resolution of each inversion (i.e. the 
horizontal FWHM of the target function) was $10\,\rm{Mm}$ for all 
depths. The vertical resolution of the inversions is $1.4\,\rm{Mm}$ for the target depth of
$1\,\rm{Mm}$, and $2\,\rm{Mm}$ for the deeper target 
depths. We find that the inversions are able to recover the simulation 
flows quite well, particularly in the upper $3\,\rm{Mm}$ of the domain, and 
correlation values are generally high here ($>0.8$). However, at the 
$5\,\rm{Mm}$ depth, the quality of our inversion has deteriorated 
significantly, and we not able to accurately recover flows there. Flow 
amplitudes are well reproduced at the two shallowest depths, but are 
underestimated at the $5\,\rm{Mm}$ depth by a factor of roughly $2.4$ in terms 
of ($v_x,v_y$) root-mean-square values.

\begin{figure*}[htb]
\begin{center}$
\begin{array}{ccc}
\includegraphics[width=0.295\linewidth,clip=]{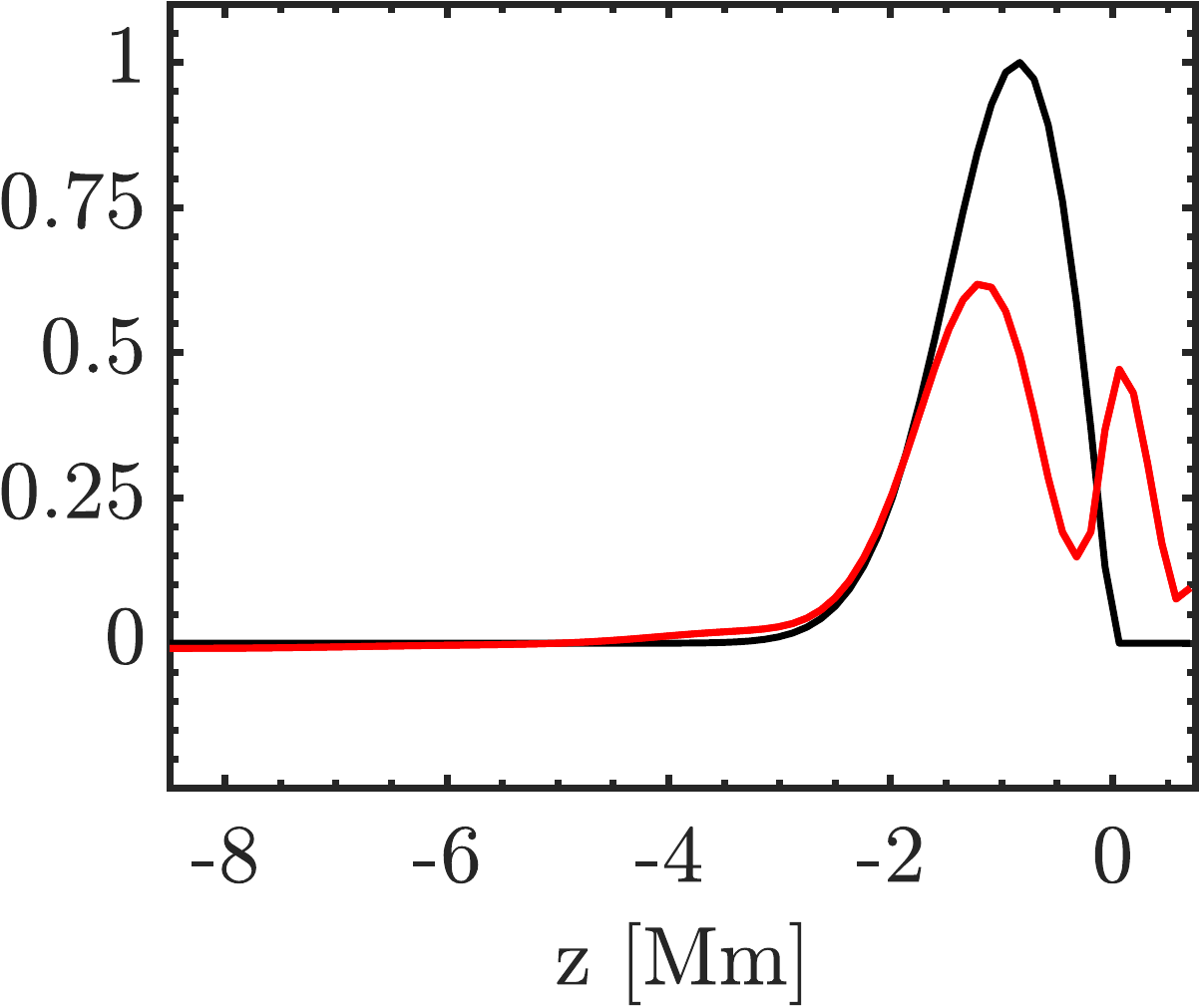}& 
\includegraphics[width=0.28\linewidth,clip=]{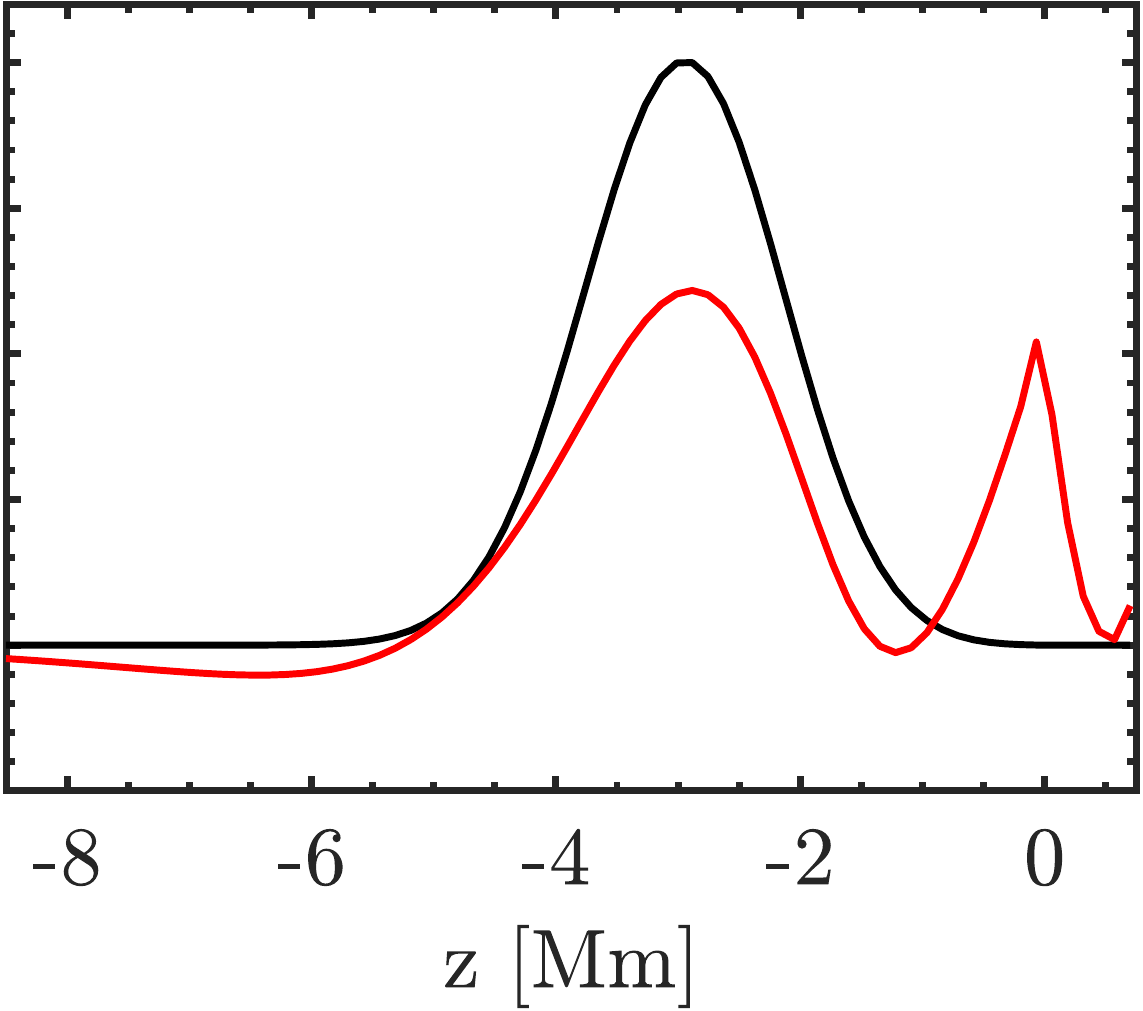}&
\includegraphics[width=0.28\linewidth,clip=]{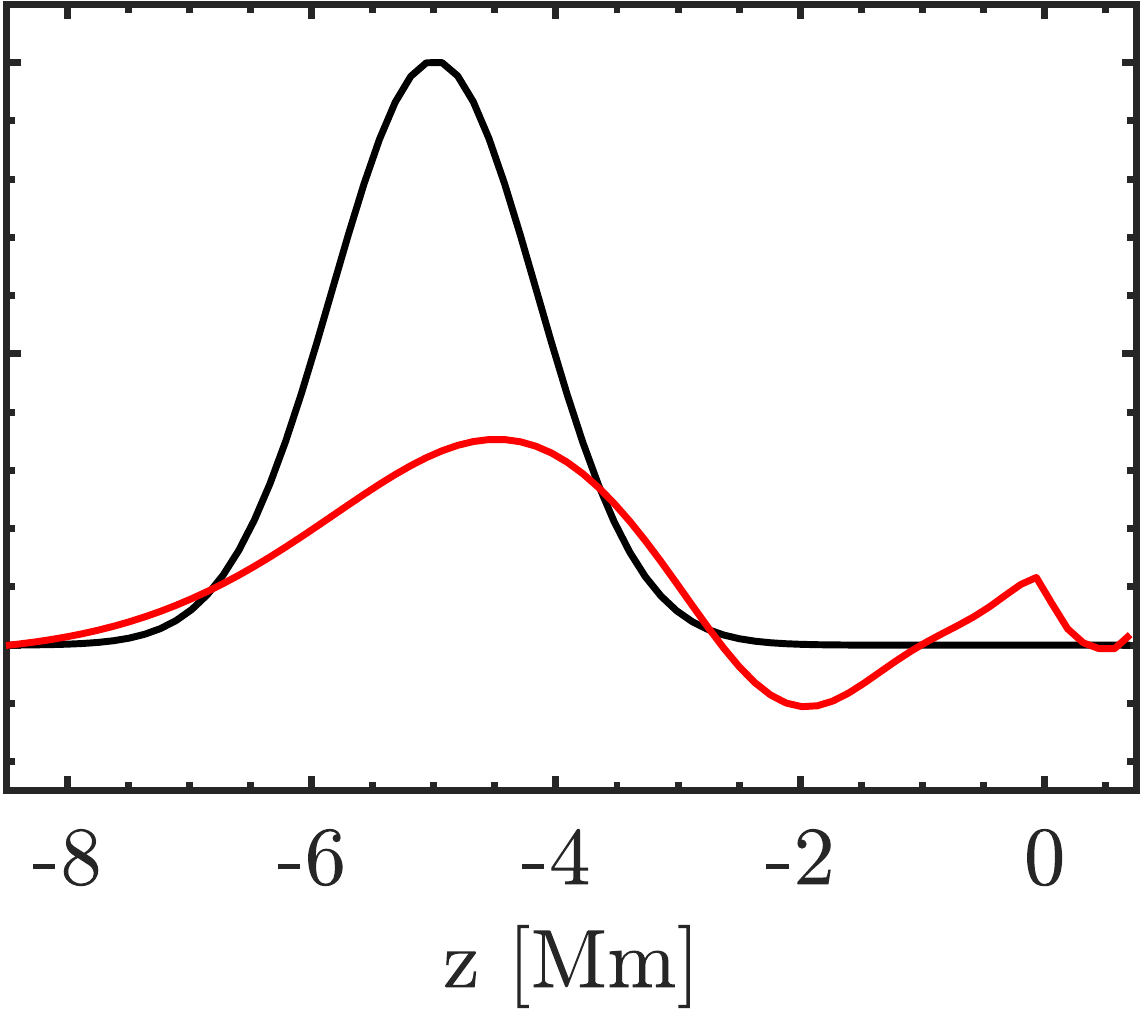}
\end{array}$
\end{center}
\caption{One-dimensional cuts along $y=x=0$ through the 
averaging kernels $\mathcal{K}_x^{x}$ (red curves) for the $1\,\rm{Mm}$ (left column), 
$3\,\rm{Mm}$ (middle column), and $5\,\rm{Mm}$ (right column) depth inversions. 
The black curves show one-dimensional cuts through the inversion target 
functions at each depth. The target functions are three-dimensional Gaussians
although, for the $1\,\rm{Mm}$ target depth, the function is multiplied by a factor 
proportional to the depth and set to $0$ for $z > 0$. 
The FWHM of the target function in the $z$-direction 
is $1.4\,\rm{Mm}$ for the target depth of $1\,\rm{Mm}$.
For the other depths the FWHM is $2\,\rm{Mm}$.
The plots are scaled such that the peak value of each target function is one.}
\label{xtarg}
\end{figure*}

One-dimensional cuts through the $x$-component of the inversion averaging 
kernels and target functions are shown in Figure~\ref{xtarg}. These figures 
show the depths targeted in the inversions along with the depths that have 
actually been sampled. The averaging kernels show some undesirable 
characteristics, most notably the large misfit observed at $z>-1\,\rm{Mm}$ for 
the $1\,\rm{Mm}$ depth (left panel), the strong near-surface contribution at 
$z>-1.5\,\rm{Mm}$ for the $3\,\rm{Mm}$ depth (middle panel), and the negative 
near-surface lobe for the $5\,\rm{Mm}$ depth (right panel). The unwanted 
near-surface sensitivity, particularly at the $3$ and $5\,\rm{Mm}$ depths, 
appears to have little effect on the recovered flows shown in 
Figure~\ref{qsvx}. This is likely, at least to some degree, due to the fact 
that the simulation flow field does not vary rapidly with depth. For example, 
the large-scale convective features in the simulation are very extended in 
depth with flows that do not show any sudden reversal in sign (e.g.\ outflow to 
inflow) or large changes in amplitude in the near-surface layers. If that were 
not the case, or if the true flow structure were not known \textit{a priori}, 
these misfit issues could make it difficult to properly interpret the results, 
particularly for inversions in the deeper layers where flows are more difficult 
to retrieve.

\begin{figure*}[htb]
\begin{center}$
\begin{array}{ccc}
\includegraphics[width=0.3699\linewidth,clip=]{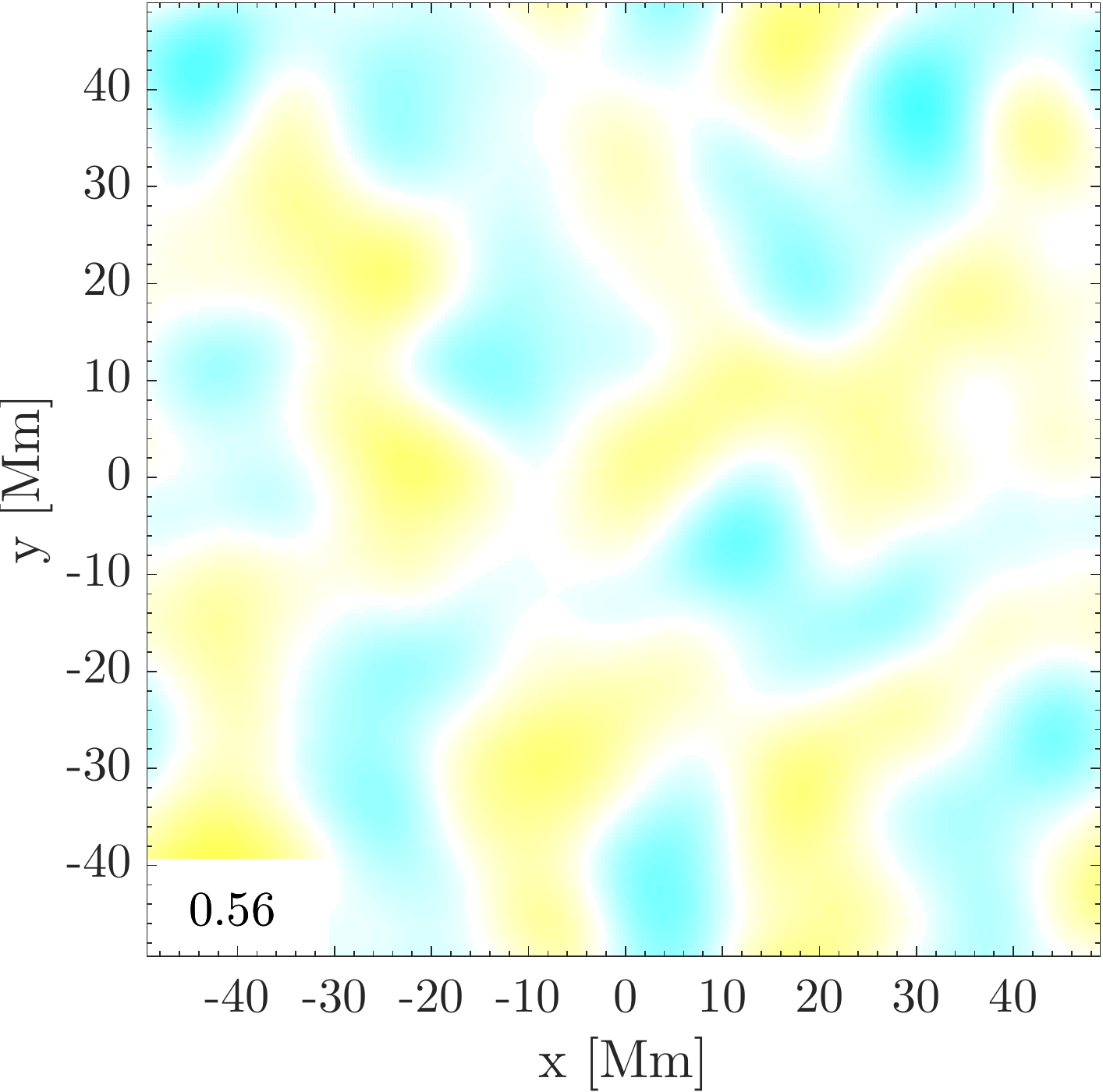}&
\includegraphics[width=0.3969\linewidth,clip=]{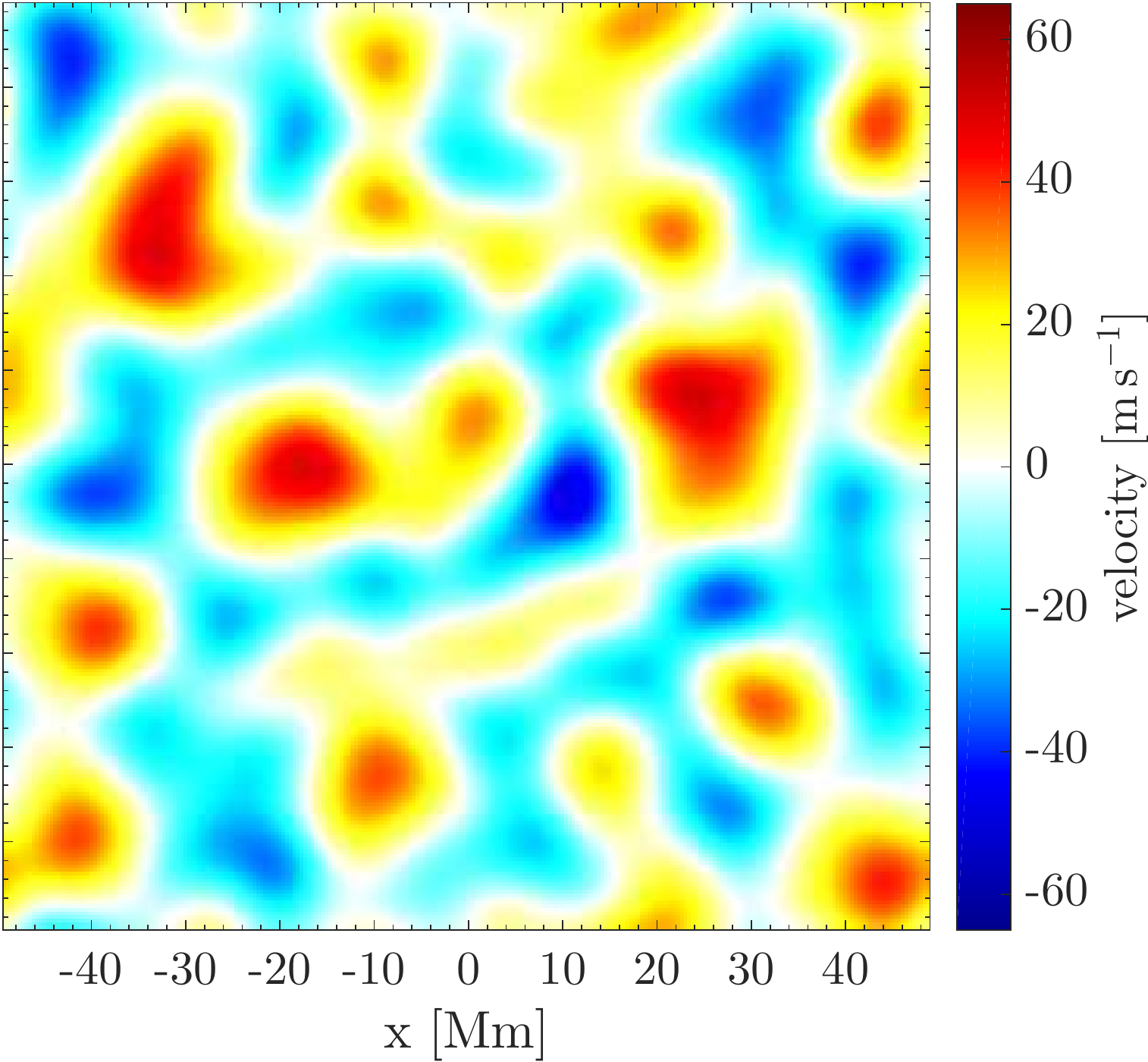}
\end{array}$
\end{center}
\caption{Flows recovered from the $1\,\rm{Mm}$ depth $v_z$ 
inversion (left) along with the target simulation flows (right). The correlation 
value between the inversion and target simulation flows is 0.56.}
\label{qsvz}
\end{figure*}

\begin{figure}[htb]
\begin{center}$
\begin{array}{c}
\includegraphics[width=1.0\linewidth,clip=]{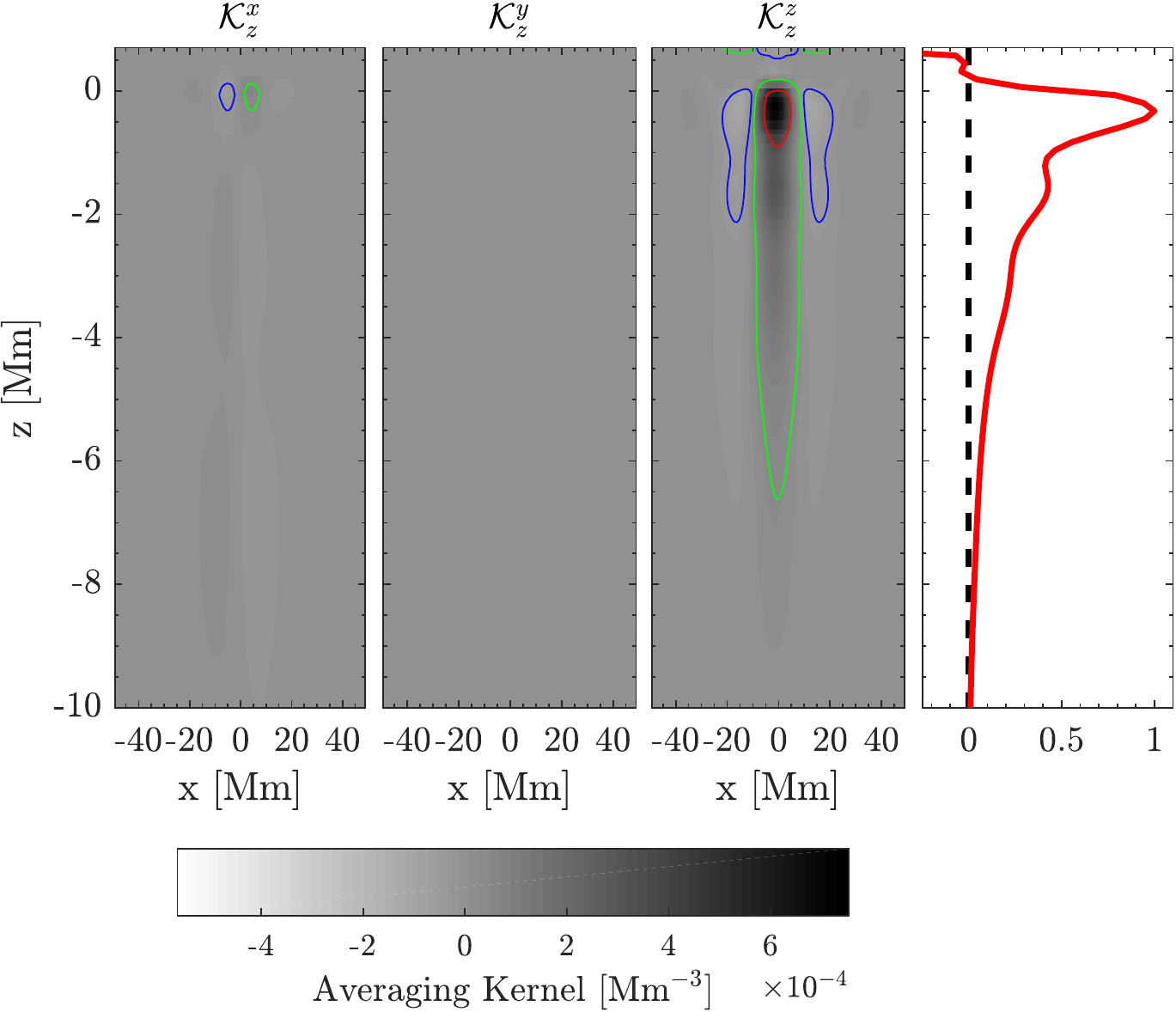}
\end{array}$
\end{center}
\caption{Cuts in depth at $y=0$ through the $v_z$ inversion averaging kernel components. 
The red contour marks the half maximum of the kernel, while the green and blue 
curves mark the $\pm0.5\,\%$ levels respectively. The right-most panel shows a 
(normalized) one-dimensional cut (red curve) along $y=x=0$ through the 
$\mathcal{K}_z^{z}$ component.} 
\label{avgvz}
\end{figure}

\subsection{Vertical Flow Inversions} \label{sec.vinvs}
An inversion was also carried out to retrieve the vertical flow component, 
$v_z$ at a depth of $1\,\rm{Mm}$. Following \citet{Svanda2013}, this inversion 
employs only kernels computed for the OI measurements. We neglect the WE and NS 
measurements as they are relatively insensitive to vertical flows and can 
actually adversely effect the recovered flows by contributing additional noise. 
The recovered flow map is shown in Figure~\ref{qsvz}. The noise level for these 
inversion is $5\,\rm{m\,s^{-1}}$, and the spatial resolution is the same as for 
the $1\,\rm{Mm}$ $v_x$ case. The two-dimensional Pearson correlation value 
between the inversion and target simulation flows is 0.56. 
Figure~\ref{avgvz} shows cuts in depth through 
the averaging kernel components for the inversion. It appears that the 
inversion is able to minimize the cross-talk well, with the 
$\mathcal{K}_z^{x}$ and $\mathcal{K}_z^{y}$ cross terms accounting for 
about $10\,\%$ or less of the maximum kernel amplitude. The averaging kernel is 
strongly peaked near the surface of the simulation domain, but shows some 
extended sensitivity over a range of depths not actually targeted in the 
inversion. The $\mathcal{K}_z^{z}$ component also exhibits broad negative side-lobes 
which we are not able to minimize effectively in the inversion.

\begin{figure*}[htb]
\begin{center}$
\begin{array}{c}
\includegraphics[width=0.98\linewidth,clip=]{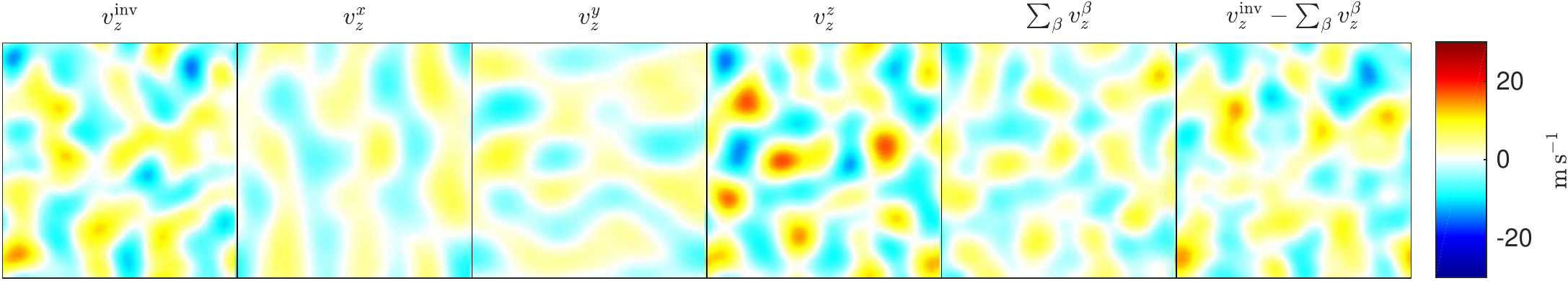}
\end{array}$
\end{center}
\caption{Noise and cross-talk contributions to $v_z$ inversions for a target depth of 1 Mm. 
From left to 
right, column $1$ shows the flows recovered from the inversions. This panel is 
identical to the map shown in Figure~\ref{qsvz}. Columns $2$\,--\,$4$ show the 
individual $v_{\alpha}^{\beta}$ terms, and column $5$ shows their sum. Column 
$6$ shows residual of columns $1$ and $5$, and is indicative of the level of 
noise which has propagated through the inversion to the solution.}
\label{vzterms}
\end{figure*}

Figure~\ref{vzterms} shows the $v_z$ inversion flow map (left-most column) 
along with the contributions from the individual terms in Equation~\ref{kv} 
(i.e., a convolution of the Figure~\ref{avgvz} averaging kernel components with 
the true simulation flows). We see again that the cross-talk minimization of 
$v_z^{x}$ and $v_z^{y}$ is reasonably effective, but somewhat less so than 
Figure~\ref{avgvz} indicates due to the strong horizontal flows with which 
$\mathcal{K}_z^{x}$ and $\mathcal{K}_z^{y}$ have been convolved. These 
cross terms have root-mean-square amplitudes that are about $50\,\%$ that of 
$v_z^{z}$. The sum of columns $2$\,--\,$4$ is shown in column $5$ and closely 
resembles the target simulation flows. This suggests that the cross-talk is not 
detrimental to the inversions in terms of being able to reproduce the correct 
flow structure of the simulation.

Column $6$ shows the residual of columns $1$ and $5$, and represents the 
contribution of noise to the recovered flows \citep{Jackiewicz2012}. Comparing 
these three columns shows that the vertical flow inversions are clearly 
dominated by this noise. Though we fail to adequately recover the vertical flow 
component here, it is important to note that these results are not dissimilar 
from the vertical flow inversions by \citet{DeGrave2014a} using kernels 
computed under the \citet{Birch2007a} prescription.

\section{Discussion}\label{sec.discuss}

We have introduced and successfully tested a set of 
sensitivity kernels for use in local helioseismology by employing them in a 
series of forward and inverse modeling comparisons using helioseismic 
holography measurements. Measured travel times computed in the lateral vantage 
compared favorably with forward-modeled ones predicted by the kernels, both in 
terms of spatial distribution and amplitude. Inversions for the horizontal flow 
components ($v_x,v_y$) employing the kernels were successful in recovering the 
simulation flow field from the upper $3\,\rm{Mm}$ of the domain, and flow 
amplitudes agreed well with those of the target flows. However, inversions 
carried out at a depth of $5\,\rm{Mm}$ were less successful in reproducing the 
flows than in the near-surface layers, and flow amplitudes were underestimated 
there by a factor of roughly $2.4$. It is important to note, though, that the 
inability of the inversions to recover flows at this depth is more a 
consequence of noise issues rather than a problem with the kernels themselves. 
A near-surface inversion for the vertical flow component failed to adequately retrieve the 
simulation $v_z$ flows. Though recovered flows correlated reasonably well 
with the target simulation flows, amplitudes were not well reproduced, and the 
inversion was dominated by noise.

This work represents the first comprehensive test of the ability of helioseismic
holography, as employed in the lateral-vantage (deep-focus) configuration, to
infer subsurface flows on spatial scales on the order of, and smaller
than, supergranules. As such, it extends and confirms the general findings of
a prior validation study \citep{Braun2007} carried out for lateral-vantage HH, but using only
forward model comparisons. Specifically, \cite{Braun2007} concluded from
a consideration of signal-to-noise (and excluding inversion-related issues) that 
supergranule-sized flows are undetectable below about 5 Mm using data spanning less than
the lifetime of a typical supergranule. The results found here also complement 
the inverse-modeling validation study performed using surface-focus holography 
\citep{Dombroski2013}, which employed regularized-least-squares (RLS)
inversions of travel-times measured from simulations of an idealized 
supergranule-like flow. 

Our results are also consistent with findings from validation studies of 
time-distance helioseismology \citep[e.g.][]{Zhao2007,Svanda2011,DeGrave2014a}.
It is now readily apparent that methods which explicitly include the minimization of
cross-talk effects such as presented here \citep[and, e.g.][]{Svanda2011,DeGrave2014a} 
offer distinct improvements
in the determination of vertical flows over methods which do not, such as the RLS
inversion of \cite{Dombroski2013}, or inversions
based on ray theory \citep{Zhao2007}. 

The dominance of realization noise for
target depths below a few Mm has lead to statistical approaches to inferring
deeper flows. A notable example is the averaging of measurements made with
respect to thousands of supergranules \citep[e.g.][]{Svanda2012,Duvall2013,Duvall2014}.
Even so, inverse modeling of these and other data apparently remaining challenging 
\citep[e.g.][]{Svanda2015,Bhattacharya2017}. 

\acknowledgements
This work is supported by NASA Heliophysics Division
through its 
Heliophysics Supporting Research (grant 80NSSC18K0066) and
Guest Investigator (grant 80NSSC18K0068) 
programs, and by the Solar 
Terrestrial program of the National Science Foundation (grant AGS-1623844).
Resources supporting this work were provided by the NASA High-End Computing 
(HEC) Program through the NASA Advanced Supercomputing (NAS) Division at Ames Research Center. 
KD acknowledges helpful discussions with Michal {\v S}vanda.

\appendix
\section{A model power spectrum}\label{sec.model}

A first step in computing flow kernels for the synthetic data is to obtain a
model for the linewidths and an empirical source function corresponding to
these data.  We begin from the azimuthally averaged power spectrum from the
simulation (30~hours total).  We fit the power spectrum of the synthetic data
with a function of the form of Equation~(53) from \citet{Birch2004}, with the
following assumptions: (1) the source correlation time is zero, this factor
will be accounted for in the empirical source function, (2) instead of assuming
a particular source depth we instead allow the source function (here denoted
$s_n(k)$) to be a free function in the fit, (3) we allow the mode frequencies
to deviate from the model~S \citep{JCD1996} frequencies, (4) the damping rates
$\gamma_n(k)$ are also treated as free parameters.  We carry out the fit in the
range $0.1$~rad~Mm$^{-1}$ $< k < 1.8$~rad~Mm$^{-1}$, using the range where
$\omega/2\pi$ is between 2.5~mHz and 5.5~mHz and the horizontal phase speed is
less then 60~km/s.  To stabilize the fit, we parameterize the damping rates at
each radial order and the real and imaginary parts of the source function at
each radial order, and the deviation from the model~S mode frequencies at each
radial order as sums of b-splines that are functions of horizontal
wavenumber.  We choose the number of b-splines for each radial order and
physical quantity by hand, with the qualitative goal of a keeping the number of
free parameters as small as possible while capturing the significant variations
of the power spectrum of the synthetic data.  In a few cases, the use of
b-splines provides more freedom than is needed; in these cases we used linear
functions of $k$. Table~\ref{table.bspline_info} shows details of the choice of
b-splines (or linear functions) for the fit.  At each radial order, the knots
of the b-splines are equally spaced between $k_{\rm min}$ and $k_{\rm max}$ for
that radial order, with duplicate knots at the end points of the interval.  The
damping rates for radial orders of four or less are fit with functions of the
form $\gamma_n(k) = \Gamma_{n,0} + \Gamma_{n,1} k^\alpha$ with $\Gamma_{n,0}$,
$\Gamma_{n,1}$, and $\alpha$ as free parameters and with a constant linewidth
is used for $n=5$ and $n=6$.

\begin{table}[ht!]
  \begin{center}
    \begin{tabular}{c|c|c|c|c|c}
      \textbf{radial order} & $k_{\rm min}$ & $k_{\rm max}$ &Re~$s$ & Im~$s$ &
$\delta\omega$ \\
      \hline
      0 & 0.9 & 1.8 & 5 & 5 & 2  \\
      1 & 0.5 & 1.8 & 5 & 5 & 4  \\
      2 & 0.3 & 1.8 & 6 & 6 & 4  \\
      3 & 0.3 & 1.8 & 5 & 5 & 4  \\
      4 & 0.3 & 1.5 & 5 & 5 & 4  \\
      5 & 0.4 & 1.2 & 3 & 3 & 2  \\
      6 & 0.5 & 1.0 & 2 & 2 & 2
    \end{tabular}
  \end{center}
    \caption{ Specification of the fitting range in $k$ ($k_{\rm min} \leq k \leq k_{\rm max}$),  the number of b-splines for the real and imaginary parts
of the source function (fourth and fifth columns), and the perturbation to mode frequency 
(last column).  
The splines are 3rd order b-splines for all cases except where only two splines are used.  In these cases, 2nd order (piecewise linear) splines are used.
 \label{table.bspline_info} }
\end{table}
Though we allow the mode frequencies to vary in the fit, we use model~S
stratification and eigenfunctions in the calculations of the Born-approximation
kernels (\S\ref{sec.kernels}).  Thus the kernel calculations do not account for
the difference between the mode frequencies in model~S and the mode frequencies
in the synthetic data.  This difference between the mode frequencies
corresponds on changes in the horizontal phases speeds of order 2\,\% and thus
presumably an error in the kernels of the same order.  Errors in the kernels of
this amplitude imply errors in the forward-modeled travel times of less than one second,
which is  well below the error estimates for the travel times (Table~\ref{tbl-lvfm}). 
We thus expect that the
errors in the inversions caused by the assumption of model S stratification
will be small compared to the errors caused by noise.

\begin{figure}[htb]
\includegraphics[width=0.9\linewidth]{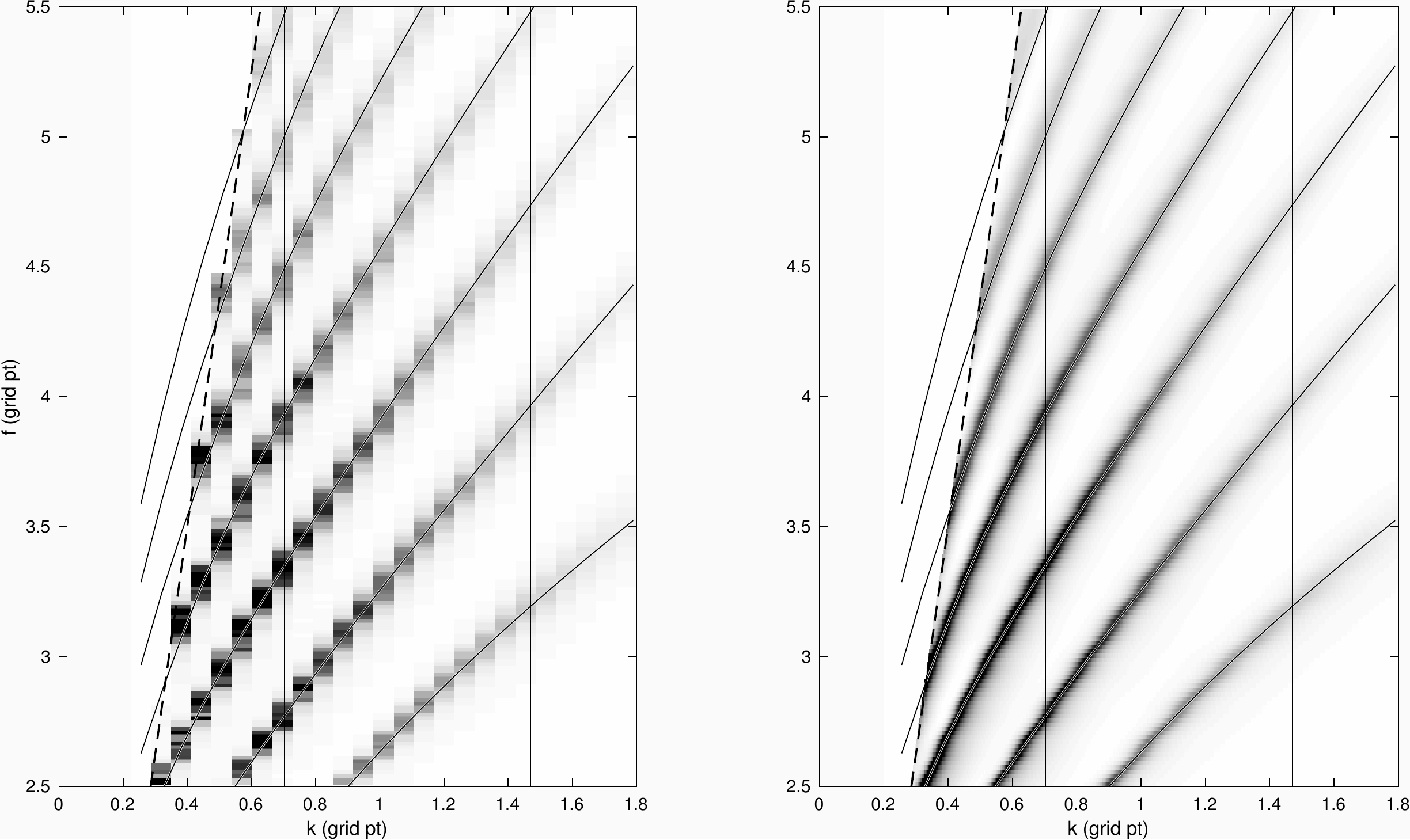}
\caption{Azimuthally averaged power spectra from 30~hours of simulation data
(left) and the resulting model power spectrum based on the fitted source
function and damping rates, but with model~S mode frequencies (right).  The
vertical lines show the locations of the cuts shown in
Figure~\ref{fig.power_cuts}. \label{fig.power_2d}}
\end{figure}

Figure~\ref{fig.power_2d} compares the model power spectrum (left) and the
model power spectrum associated with the kernel calculation (right).  For the
part of the diagram with horizontal phase speed less then 55~km/s, there is
reasonable qualitative agreement.

\begin{figure}[htb]
\includegraphics[width=0.45\linewidth]{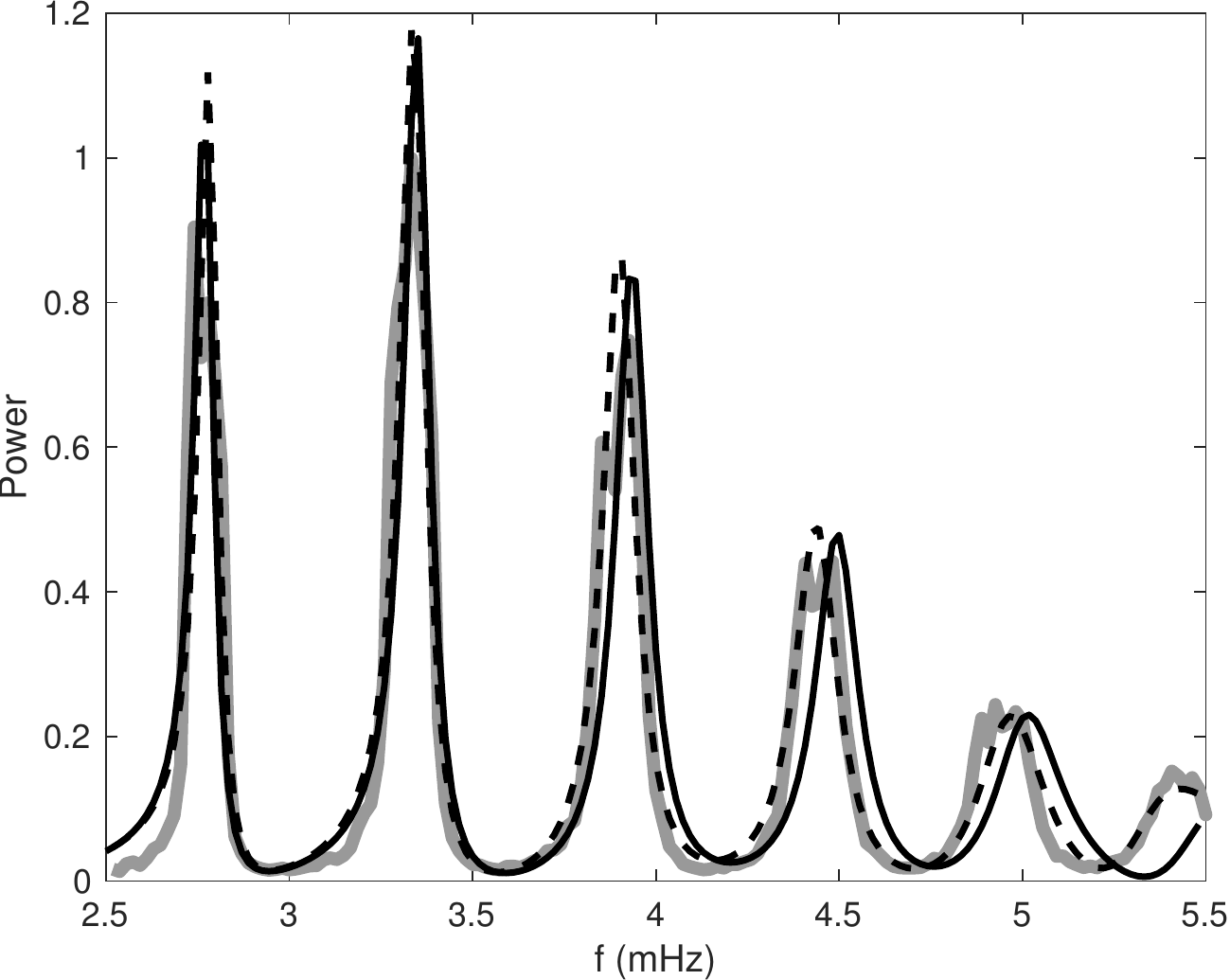}
\includegraphics[width=0.45\linewidth]{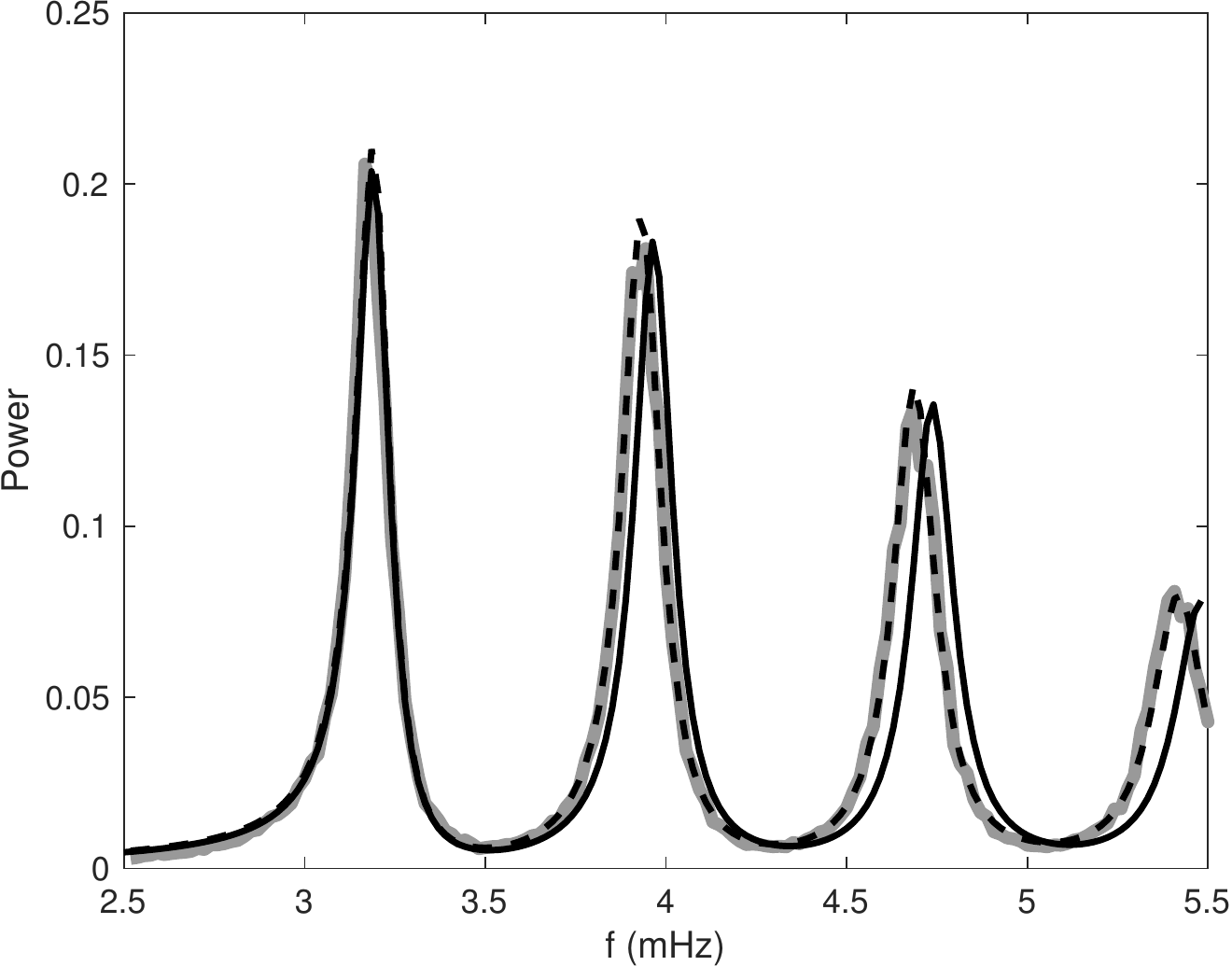}
\caption{Slices through the power spectra from Figure~\ref{fig.power_2d} at
$\ell=489$ and $\ell=1023$.  The simulation power spectrum is shown in the
thick gray line, the fit is the dashed black line, and the model power spectrum
based on the fitted source function and linewidths, but with model~S mode
frequencies is shown in the solid black line. \label{fig.power_cuts}}
\end{figure}

Figure~\ref{fig.power_cuts} shows two example slices through the power spectra
at constant horizontal wavenumber.  In these slices, the result of the fit is
shown along with the power spectra from the synthetic data and the model power
spectrum from the kernel calculation.  As discussed earlier, the kernel
calculation is carried out using the model~S stratification, and so the
resonance frequencies are slightly different than in the simulations.

\section{Flow kernels}\label{sec.kernels}

For the calculation of the kernels, we work in a coordinate system where ${\bf 
r}$ is horizontal position and $z$ is height measured from the photosphere 
($z=0$).  The background model is translation invariant and given by a 
plane-parallel version of model~S. 

At each temporal frequency, the ingression $H_-$ at horizontal position $\br$ 
for a particular focus height $z$ is related to the observed wavefield $\phi$ by
\begin{equation}
H_-(\br) = \int G_-^P(\br'-\br)\phi(\br') \;  d \br'
\label{eq.ingression}
\end{equation}
where $G_-^P$ is the anti-causal Green's function multiplied by the appropriate 
pupil function $P$ and data-analysis filter function.
The Green's function, the pupil function, and the filter 
function all depend on the focus depth (see \S\ref{sec.ttmeas}).  For the sake of 
readability, we have also suppressed the notation showing that $H_-$, $G_-$, 
and $\phi$ are all functions of temporal frequency.  The integral is taken over 
all horizontal positions where the pupil function is not zero. 
Equation~(\ref{eq.ingression}) shows that the ingression is the result of 
filtering the wavefield with a non-axisymmetric filter (the 2D Fourier 
transform of $G_-^P$).   The egression is related to the wavefield by an 
analogous equation, but with $G_-$ replaced by the causal Green's function 
$G_+$, and the pupil function replaces by its appropriate counterpart (e.g, for 
the NS travel-time difference, if $P$ is the north pupil then $P'$ 
is the south pupil). 

At each temporal frequency, The lateral-vantage ingression-egression covariance 
$C$ at the horizontal position $\br={\bf 0}$ is 
\begin{equation}
C(\br) = H^*_-(\br) H_+(\br) \; .
\label{eq.full_C}
\end{equation}
The ingression-egression covariance is a time-distance covariance at zero 
distance (recalling the ingression and egression are filtered versions of the 
wavefield). We can use equations~(11)-(13), which already allow for arbitrary 
non-axisymmetric filters, from \citet{Birch2007a} to compute the linear 
sensitivity of $C$ to flows. 

Travel-time shifts are measured from the ingression-egression covariance $C$ as
\begin{equation}
\tau = {\rm Arg}\left[ \bar{C} \right] / \bar{\omega} \; ,
\end{equation}
where $\bar{C}$ is the sum of $C$ over all frequency bins and $\bar{\omega}$ is 
the average frequency weighted by $| C |$.  The perturbation 
$\delta\tau$ to the travel-time shift coming from a perturbation 
$\delta\bar{C}$ to the frequency-summed ingression-egression covariance 
is:
\begin{equation}
\delta \tau = {\rm Im}[ \bar{C}^* \delta \bar{C}] / \left[ 
|\bar{C}|^2\bar{\omega} \right] \; .
\end{equation}
This equations shows that the linear sensitivity of the travel-time shift to 
flows is a linear combination of the kernels for the ingression-egression 
covariances (Eq.~(\ref{eq.full_C}).  The resulting vector-valued kernels ${\bf 
K}$ satisfy:
\begin{equation}
\delta\tau(\br)  = \int\!\!\int\!\!\int {\bf K}(\br'-\br,z) \cdot {\bf 
u}(\br',z) \; d \br' d z \; .
\end{equation}

\section{Example kernels}

\begin{figure}[htb]
\includegraphics[width=0.9\linewidth]{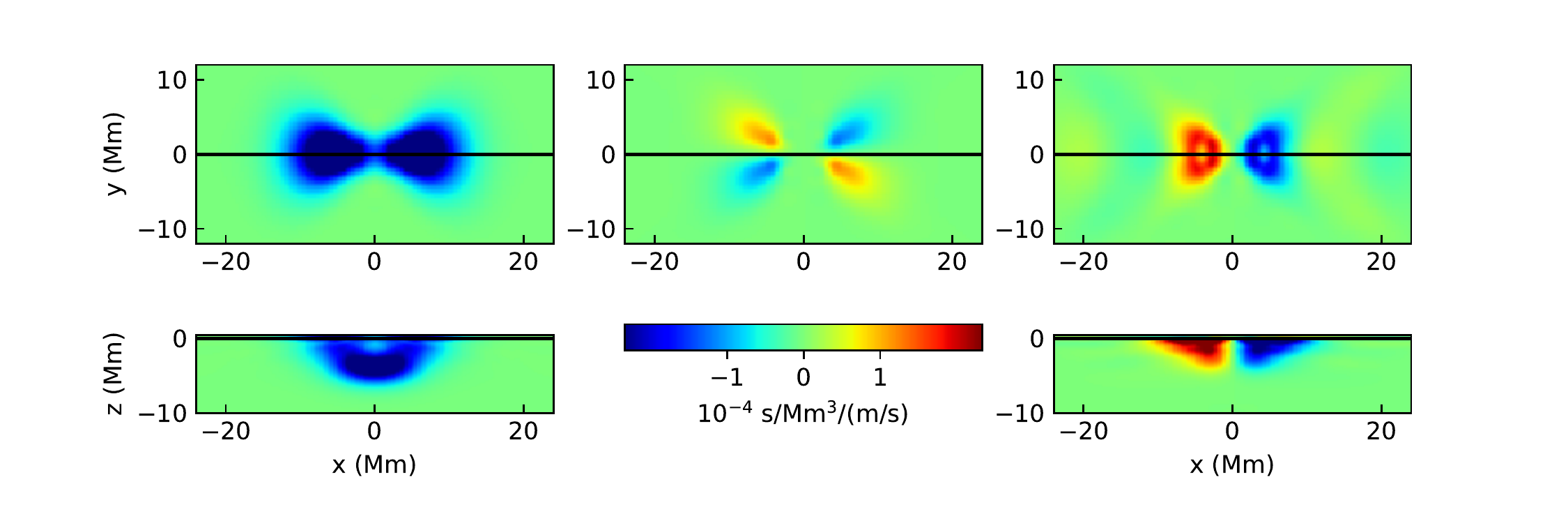}
\caption{Slices through the kernels $K_\beta^\mathrm{we}$ for a focus depth of 3.97~Mm.  
These kernels provide the sensitivity of the travel-time difference in the $x$ 
direction to arbitrary 3D flows.  The left (middle, right) column shows slices 
through the $K_x^\mathrm{we}$ ($K_y^\mathrm{we}$, $K_z^\mathrm{we}$) kernel.  The top row shows slices at the 
photosphere of the model and the second row shows vertical slices at $y=0$.  
For the case of a horizontally uniform flow, the travel-time shift would depend 
only on the $K_x^\mathrm{we}$ kernel.\label{fig.kernel_x}}
\end{figure}

\begin{figure}[htb]
\epsscale{1.0}
\plotone{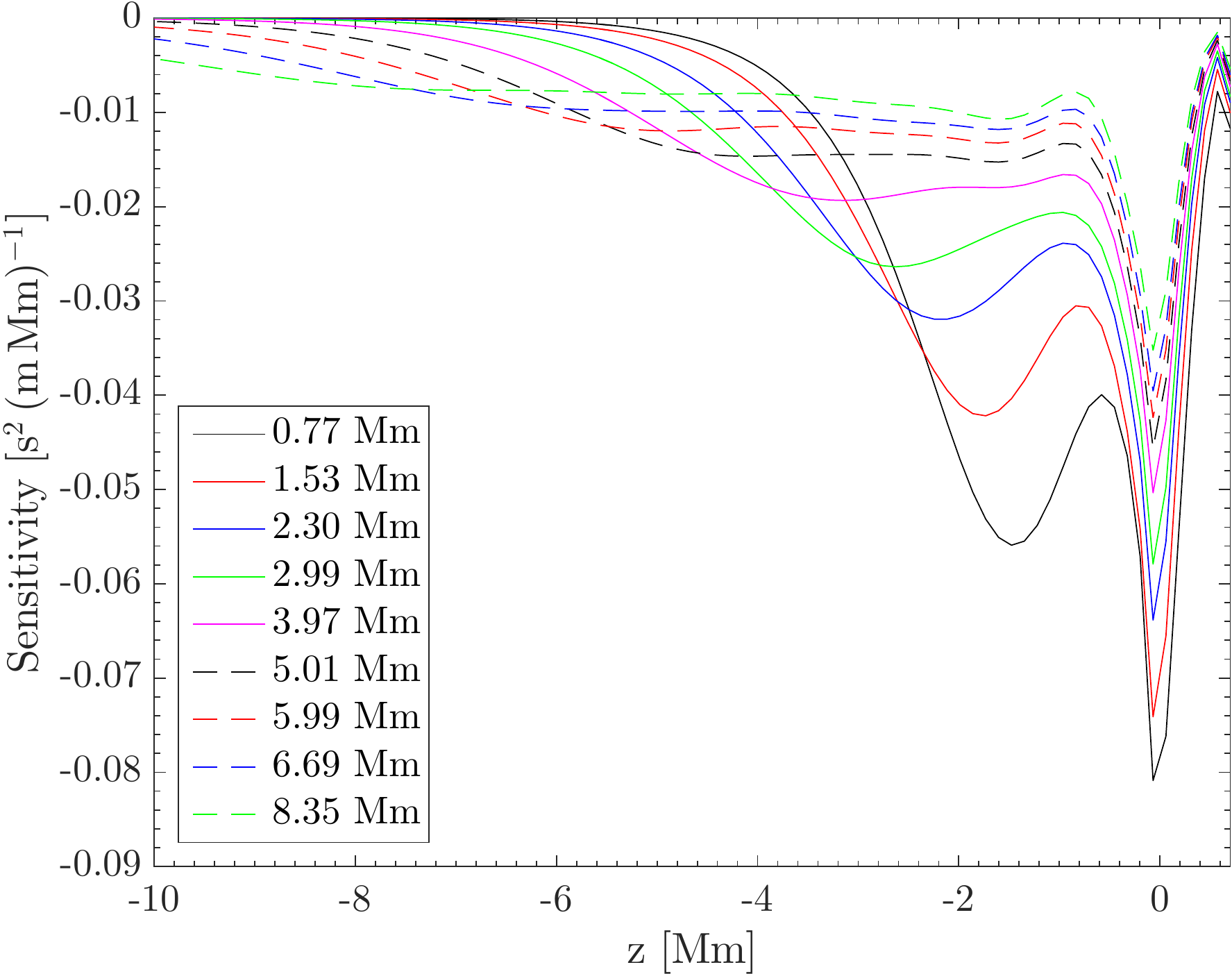}
\caption{The depth dependence of the horizontally-integrated kernels $K_x^\mathrm{we}$ for
the different focus depths as indicated.
}
\label{fig.lvkerns}
\end{figure}

Figure~\ref{fig.kernel_x} shows slices through the kernels $K_\beta^\mathrm{we}$ for the 
focus depth 3.97~Mm. The kernel $K_x^\mathrm{we}$ is symmetric in both $x$ and $y$, the 
kernel $K_y^\mathrm{we}$ is anti-symmetric in both $x$ and $y$, and the kernel $K_z^\mathrm{we}$ is 
anti-symmetric in $x$ and symmetric in $y$.  The $K_x^\mathrm{we}$ and $K_z^\mathrm{we}$ kernels 
have larger amplitudes than the $K_y^\mathrm{we}$ kernel; as expected travel-time 
differences in the $x$ direction are mostly sensitive to flows in the $x$ 
direction and vertical flows. The depth dependence of the $K_x^\mathrm{we}$ is shown in 
more detail in Figure~\ref{fig.lvkerns}.

\begin{figure}[htb]
\includegraphics[width=0.9\linewidth]{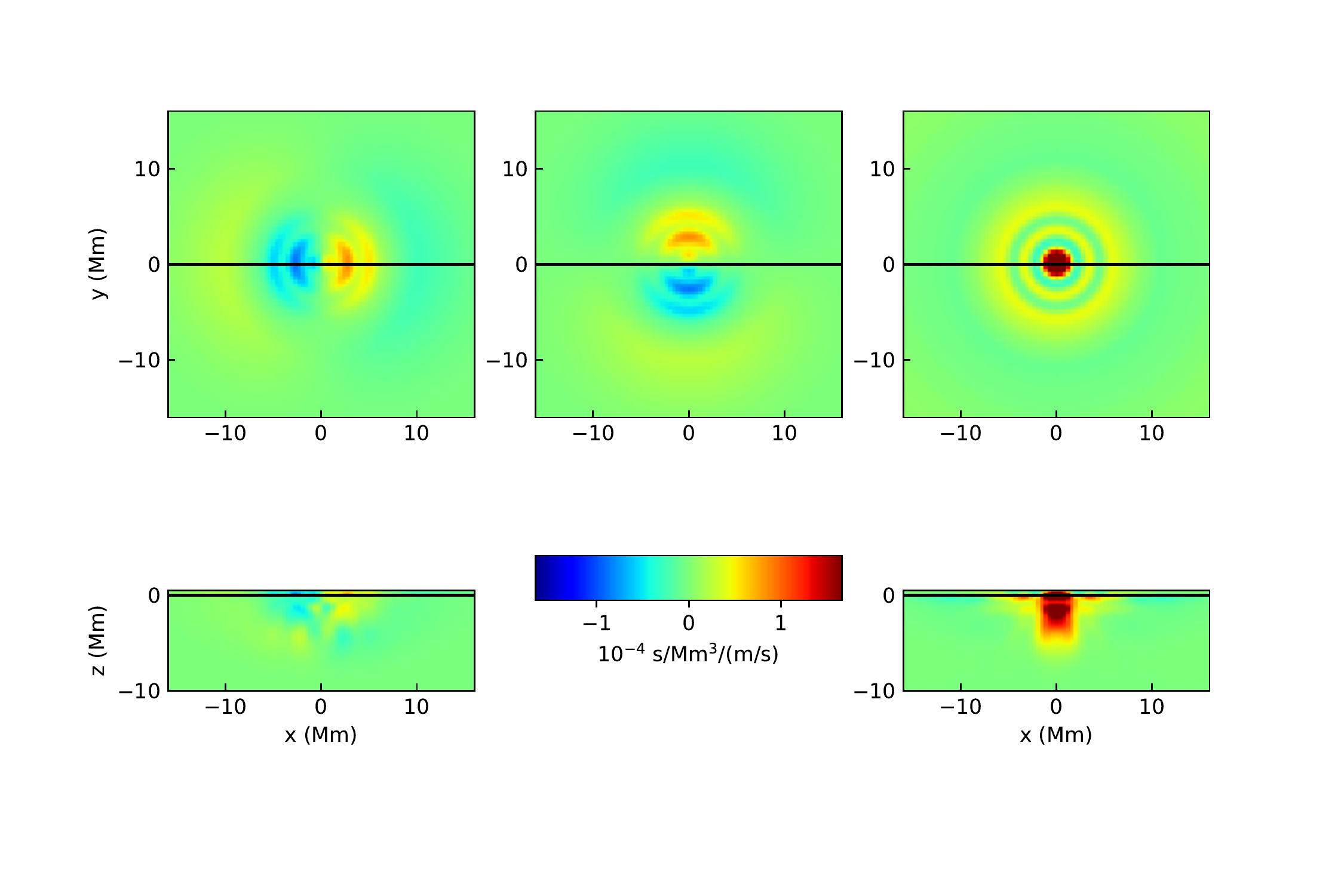}
\caption{Slices through the kernels $K_\beta^\mathrm{oi}$ for the sensitivity of the 
OI travel-time difference for the focus depth of 3.97~Mm.  The layout is the 
same as in Figure~\ref{fig.kernel_x}.  The symmetries are different than in the 
case of the travel-time difference in the $x$ direction.  In the case of OI 
measurements, the travel-time sensitivity is dominated by sensitivity to 
vertical flows near the horizontal focus point.\label{fig.kernel_oi}}
\end{figure}

Figure~\ref{fig.kernel_oi} shows slices through the kernels $K_\beta^{\rm oi}$ 
for the focus depth 3.97~Mm.  The symmetries are different than for the case of 
travel-time difference in the $x$ direction.  The kernel $K_x^\mathrm{oi}$ is 
anti-symmetric in $x$ and symmetric in $y$.  The kernel $K_y^\mathrm{oi}$ is symmetric 
in $x$ and anti-symmetric in $y$.  The kernel $K_z^\mathrm{oi}$ is cylindrically 
symmetric about the $z$ axis.  The largest sensitivity is to vertical flows 
near the axis.  The sign is such that up-flows cause an increase in the OI
time difference.

\bibliographystyle{/export/home/dbraun/Macros/apj}
\bibliography{/export/home/dbraun/Macros/db}

\end{document}